\documentclass[secnumarabic, graphics,floatfix, nofootinbib,tightenlines,nobibnotes, aps, 12pt]{revtex4}
\usepackage{graphicx}
\usepackage[english]{babel}
\usepackage[utf8]{inputenc}
\usepackage{amsmath,amssymb}
\usepackage{amsfonts}
\usepackage{multirow}
\usepackage{booktabs}
\usepackage[section]{placeins}
\usepackage{float}%
\usepackage[colorlinks]{hyperref} 
\usepackage{graphicx,epstopdf}
\begin{document}

\newcommand{\bq}{\begin{equation}}
\newcommand{\eq}{\end{equation}}
\newcommand{\bqn}{\begin{eqnarray}}
\newcommand{\eqn}{\end{eqnarray}}
\newcommand{\nb}{\nonumber}
\newcommand{\lb}{\label}
%

\title{The initial structure of chondrule dust rims II: charged grains}

\begin{abstract}

In order to characterize the early growth of fine-grained dust rims (FGRs) that commonly surround chondrules, we simulate the growth of FGRs through direct accretion of monomers of various sizes onto the chondrule surfaces. Dust becomes charged to varying degrees in the radiative plasma environment of the solar nebula (SN), and the resulting electrostatic force alters the trajectories of colliding dust grains, influencing the structure of the dust rim as well as the time scale of rim formation. We compare the growth of FGRs in protoplanetary disks (PPD) with different turbulence strengths and plasma conditions to previous models which assumed neutral dust grains (Xiang, C., Carballido, A., Hanna, R.D., Matthews, L.S., Hyde, T.W., 2019). We use a combination of a Monte Carlo method and an N-body code to simulate the collision of dust monomers of radii 0.5 - 10 $\mu$m with chondrules whose radii are between 500 and 1000 $\mu$m: a Monte Carlo algorithm is used to randomly select dust particles that will collide with the chondrule as well as determine the elapsed time interval between collisions; at close approach, the detailed collision process is modeled using an N-body algorithm, Aggregate Builder (AB), to determine the collision outcome, as well as any restructuring of the chondrule rim. For computational expediency, we limit accretion of dust monomers to a small patch of the chondrule surface. The collisions are driven by Brownian motion and coupling to turbulent gas motion in the protoplanetary disk. The charge distribution of the dust rim is modeled, used to calculate the trajectories of dust grains, and then analyze the resulting morphology of the dust rim.  In a weakly turbulent region, the decreased relative velocity between charged particles causes small grains to be repelled from the chondrule, causing dust rims to grow more slowly and be composed of larger monomers, which results in a more porous structure. In a highly turbulent region, the presence of charge mainly affects the porosity of the rim by causing dust particles to deviate from the extremities of the rim and reducing the amount of restructuring caused by high-velocity collisions.


\end{abstract}

\maketitle
\section{Introduction}
\renewcommand{\theequation}{1.\arabic{equation}} \setcounter{equation}{0}
Chondrules are the primary constituents of chondritic meteorites, and those in carbonaceous chondrites are generally surrounded by rims of micron and submicron-sized mineral grains. Several lines of evidence, such as the difference of chemical abundances (Leitner et al. 2016; Metzler 2004, 1992) and the smaller grain size in FGRs than in the matrix (Ashworth 1977; Brearley 1993, 1999; Zolensky et al. 1993) indicate that fine-grained dust rims (FGRs) were formed by accretion of grains onto the underlying chondrule cores in a nebular setting, before the rimmed chondrules were incorporated into their parent bodies (Metzler et al. 1992, Morfill et al. 1998, Brearley et al. 1999). In addition, the linear/power law relationship between the thickness of dust rims and the diameter of the rimmed cores found in experimental observations (Metzler et al. 1992; Hanna \& Ketcham 2018) is consistent with the results from numerical simulations assuming a nebular origin (Ormel et al. 2008; Carballido 2011; Cuzzi 2004).

Although the nebular origin is the most widely accepted mechanism for FGR formation, others have proposed that FGRs were formed in the parent body environment, either by attachment and compaction of dust onto chondrules in regolith (Sears et al. 1993, Trigo-Rodriguez et al. 2006, Takayama \& Tomeoka 2012), or through pervasive aqueous alteration of chondrules (Sears et al. 1993, Takayama \& Tomeoka 2012). These theories are supported by the embayment textures along the chondrule-rim boundary (Takayama \& Tomeoka, 2012) and the low rim porosity which is inconsistent with the high-porosity structure formed in experimental simulations of preplanetary dust coagulation (Blum \& Wurm 2000, Blum \& Schrapler 2004). Beitz et al. (2013) performed experiments of impacts between chondrule analogs and different dust materials and found larger dust rim porosities than those found in chondrites, leading to the conclusion that FGRs around chondrules can not form in dynamic compaction processes. Thus, there are still open questions about what mechanisms reduce the primordial porosity of FGRs as they evolve.

Previous theoretical models of FGR formation employed various methods, including integrating an evolution equation (Morfill et al. 1998), using a semi-analytical model (Cuzzi 2004), and employing a Monte Carlo method (Ormel et al. 2008), all of which were based on experimental and theoretical results from the physics of dust collisions (Chokshi et al. 1993, Dominik \& Tielens 1997, Blum \& Schrapler 2004). Carballido (2011) furthered the approach of Morfill et al. (1998) by simulating dust sweep-up by chondrules in a local, magnetohydrodynamic (MHD) turbulent model of the solar nebula (SN). However, all these models did not resolve the actual rim structure, and assumed that the dust grains comprising the rims are electrically neutral. In our preceding paper (Xiang et al. 2018, hereafter Paper I), we followed the assumption of electrically neutral grains, and developed a molecular dynamics model to simulate collisions between dust and chondrules, taking into consideration the detailed collisional physics, such as the trajectory of the incoming particle, restructuring after the collision, and the resulting morphology of the dust rim. The main findings of Paper I are:

a.	FGRs formed in environments with strong turbulence are more compact and grow more rapidly than FGRs formed in weak turbulence.

b.	The time needed to build FGRs of a certain thickness decreases approximately linearly with the turbulence strength.

c.	FGR thickness scales linearly with chondrule radius, as in observed rims, and this linear relation becomes steeper over time. 

d.	Grain-size coarsens toward outer portions of FGRs, consistent with observations. 

e.	FGRs formed by aggregates of dust grains are more porous than those formed by addition of single spherical monomers.
\\

Dust grains in the SN can become charged to varying degrees in the radiative plasma environment. The trajectories of colliding dust grains can be altered by the electrostatic force acting between the grains, affecting their coagulation probability as well as their impact velocity (Matthews et al. 2012, 2016; Ma et al. 2013; Okuzumi et al. 2009). The electrostatic force therefore influences the structure of the dust rim (porosity, monomer size distribution, etc.), as well as the time scale of rim formation. The effect of dust charge on this process is modified by the turbulence strength, as dust grains entrained in regions with relatively strong turbulence have a greater probability of overcoming the electrostatic barrier, and hence reaching the rimmed chondrule surface. The collection of dust particles can be inhibited by the increasing Coulomb repulsion force, and the strength of the electrostatic barrier is roughly proportional to the aggregate size (Okuzumi 2009, 2011a, 2011b). For these reasons, it is important to take into account grain charge in models of FGR formation.

There are several proposed mechanisms leading to turbulence in a PPD, including the Magnetorotational Instability (MRI; Balbus \& Hawley 1998), Vertical Shear Instability (Urpin, 2003), Convective Overstability (Latter, 2015), Zombie Vortex Instability (Marcus et al. 2015), and the Streaming Instability (Raettig et al. 2015; Johansen \& Youdin 2007). The dominant process varies with disk location, determined by the cooling times defined by disk opacities. In this work, we do not specify the origin of the turbulence but characterize its strength through the dimensionless hydrodynamical viscosity parameter $\alpha$ of Shakura \& Sunyaev (1973). Similarly, we do not choose particular plasma conditions, but instead, use a range of dust surface potentials which are relevant for many regions of a protoplanetary disk. We aim to quantify the physical characteristics and timescales of FGRs over a range of nebular conditions, and investigate the interplay between the effects of charge and turbulence on FGR formation. If the nebular hypothesis is correct, the resulting FGR structure could ultimately be used to infer values of gas velocities, turbulent viscosity, and ionization state of the solar nebula (Ormel et al. 2008; Matthews et al. 2012; Okuzumi et al. 2009).

We simulate FGRs formed by direct accretion of spherical monomers of various sizes onto chondrule surfaces. As in Paper I, we assume that the chondrule accretes dust isotropically (Bland et al. 2011), and restrict our study to a small patch on the chondrule surface for computational expediency. In this model,  both the chondrule parent body and the dust grains are charged, and we include the electrostatic interactions during the rimming/accretion process.  The structure of the resulting dust rim (porosity, monomer size distribution and thickness), is examined in order to identify the manner in which charge alters the rim structure and growth rate.

This paper is organized as follows. In Section 2, we introduce the method for calculating the charge distribution on the dust rim, and describe the ``detailed Monte Carlo'' method for modeling the collection of dust on the chondrule surface, which encompasses the dust dynamics, restructuring, as well as the selection process. In Section 3, we investigate the effects of the charge, turbulence strength, and chondrule size on rim growth, and present the results from analysis of the rim structure (porosity, monomer size distribution and thickness) as well as the time to build the rims. The main conclusions are summarized in Section 4.
\section{NUMERICAL METHOD           
}
\renewcommand{\theequation}{2.\arabic{equation}} \setcounter{equation}{0}
The initial structure of FGRs is determined by interactions between micrometer-sized dust grains and mm-sized chondrules. Small particles entrained in a turbulent gas flow develop relative velocities due to the difference in their coupling times with the gas. The collision energy, a function of the particles' relative velocity, as well as the potential energy of the charged grains, determines the collision outcome: incoming dust particles can stick at the point of contact, cause restructuring of the dust rim, bounce, or be repelled if the particles' kinetic energy is not great enough to overcome the electrostatic barrier.

\begin{figure*}[!htb]
\includegraphics[width=7cm]{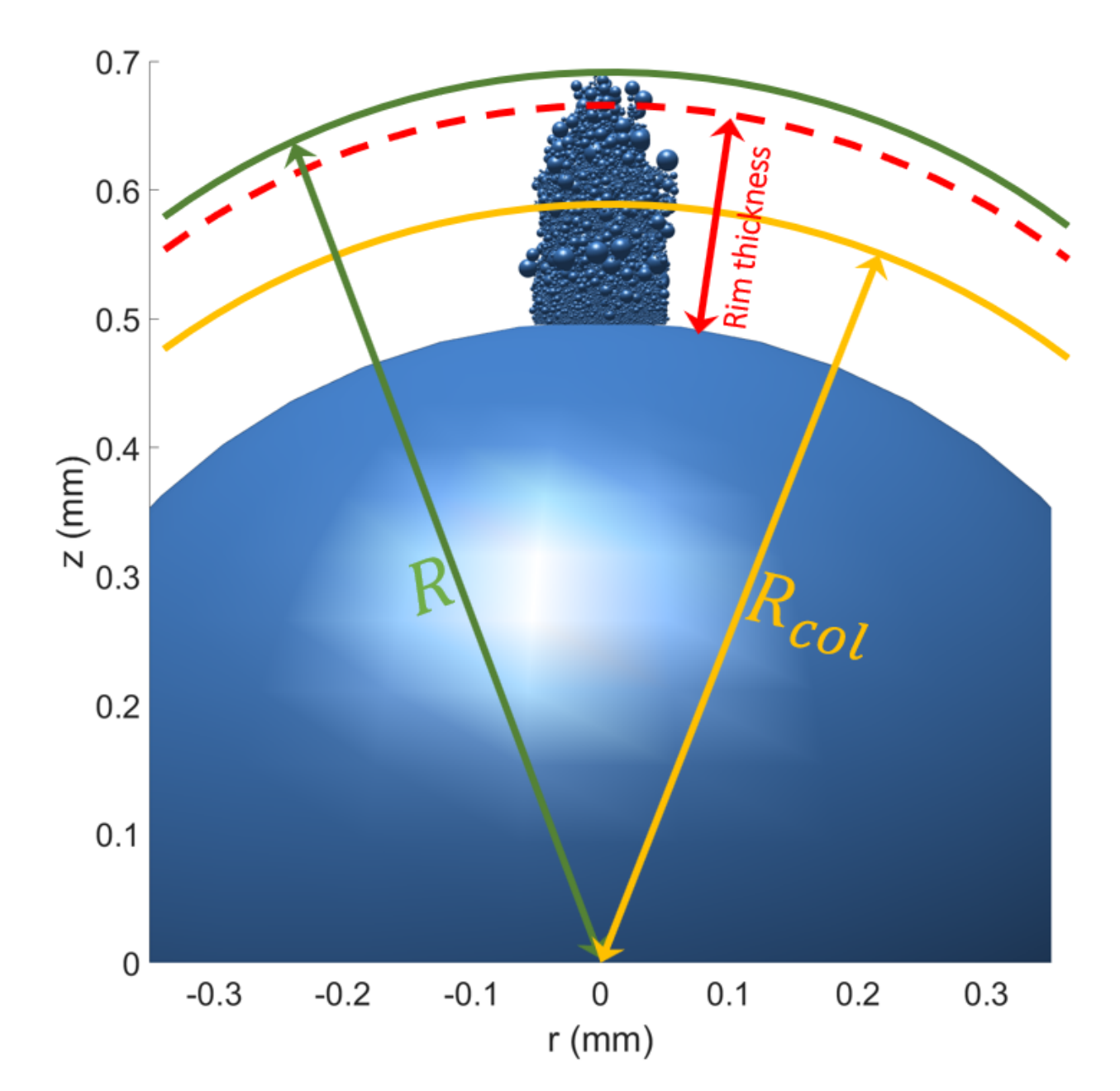} 
\caption{Illustration of physical radius $R$, collisional radius $R _{col}$, and rim thickness.}
\label{fig0}
\end{figure*}

In our treatment of chondrule rim growth, the factors that affect the coagulation process are the probability that two particles travel towards each other (determined by their cross-sectional area and relative velocity) and the type of interaction between them, which determines the collision outcome (i.e., sticking, restructuring, repulsion, etc.). We use a combination of a Monte Carlo method and an N-body code to model these two factors: a Monte Carlo algorithm is used to randomly select dust particles that will collide with the chondrule as well as determine the elapsed time interval between collisions; at close approach, the detailed collision process is modeled using an N-body algorithm, Aggregate Builder (AB), to determine the collision outcome, as well as any restructuring of the chondrule rim. Upon collision, the charge of the rimmed chondrule is recalculated using a numerical model (OML-LOS, Matthews et al. 2012) which self-consistently calculates the charge accumulated on the irregular rim surface. Thus this model represents a ``detailed-Monte Carlo' method. Each simulation begins with a (sub)-millimeter-sized spherical chondrule, with a radius between 500 $\mu$m and 1000 $\mu$m, placed at the origin. The potentially colliding dust grains are selected from a population of 10,000 dust particles with radii ranging from 0.5 to 10 $\mu$m, with a power law size distribution $n(r)dr\propto r^{-3.5}dr$, where $n(r)dr$ is the number of particles in the size interval $(r,r+dr)$ (Mathis et al. 1977). The rim growth is modeled by restricting the collection of dust to a patch on the surface of the chondrule, as illustrated in Fig. \ref{fig0}.

Two definitions of radius, collisional radius $R _{col}$ and physical radius $R$, are used in our simulation. The collisional radius $R _{col}$ is defined as the sum of the radius of the chondrule core and half of the maximum radial extent of the rim to the surface of the chondrule core (indicated by the yellow arrow in Fig. \ref{fig0}). The physical radius is the maximum radial extent from any point on the rim to the center of the chondrule core (indicated by the green arrow in Fig. \ref{fig0}). $R _{col}$ and $R$ are are used to calculate the collision rate of dust particles with the chondrule (Section \ref{sec:mc}) and to set the initial distance between chondrule and dust particle (Section \ref{sec:ab}), respectively. In order to characterize the growth of dust rims, we define the rim thickness as the distance from the surface of chondrule core encompassing 95\% of the total rim mass (indicated by the red arrow in Fig. \ref{fig0}).


\subsection{Monte Carlo collision selection}\label{sec:mc}


 The dust particles that collide with the chondrule are selected from an initial dust population based on their collision rate $ C_{ch,d}$, where $ch$ stands for chondrule and $d$ stands for dust particle (Ormel et al. 2007, 2008),
\bqn
\lb{10}
C_{ch,d}=\sigma _{ch,d}\Delta v_{ch,d}/V.
\eqn
The collision rate depends on the collision cross section $\sigma _{ch,d}= \pi(R _{col}+r_{d})^{2}$, where $R _{col}$ is the radius of the rimmed chondrule (see Fig. \ref{fig0}) and $r_{d}$ is the radius of the dust particle, the relative velocity of the dust with respect to the chondrule $\Delta v_{ch,d}$, and the volume $V$ of the simulated region, set by the density of the dust in the region.

In order to reduce the computational cost, the dust particles are grouped into 100 logarithmic bins by their radii, and the collision rates $ C_{ch,d}$ are calculated using the average radii and masses of particles in each bin. In each iteration, time is advanced by a random interval $ \tau$ based on the collision rate (Gillespie 1975),

\bqn
\lb{10}
\tau=-ln(r)/C_{tot}
\eqn
with $r$ a random number between zero and one, and $C_{tot}=\sum_{j=1}^{N}C_{ch,d}$ the sum of collision rates over all dust particles, where $N$ is the total number of dust particles.

\subsection{Detailed interaction model}\label{sec:ab}
After the potential colliding dust grain is selected by the MC algorithm, the collision outcome is modeled by Aggregate Builder (AB; Matthews et al. 2012), which takes into account the morphology of the dust on the rim, the trajectory of the incoming particle, and the electrostatic interaction.

For computational expediency, we restrict dust particles to accumulate on a circular patch 100 $\mu$m in diameter of the chondrule surface. The dust particle is shot towards a randomly selected point on the target area from a random direction (the angle between this direction and the normal to the patch at the selected point is uniformly distributed between $0^{\circ}$ and $60^{\circ}$, so that it is less likely that dust particles impact the side of the dust pile). The initial distance between the chondrule and the incoming dust particle is set to be $2.5R$, with $R$ the maximum radius of the rimmed chondrule, as shown in Fig. \ref{fig0}. The initial relative velocity between the chondrule and the dust grains is set assuming that the dust is coupled to the turbulent eddies in the gas in the protoplanetary disk, (Ormel et al. 2008) 

\bqn\label{eq:vturb}
\lb{3}
v_{T}=\
v_gRe^{1/4}( \rm{St}_{1}- St_{2})
\eqn
where $v_g$ is the gas speed, and $Re$ is the Reynolds number, defined as the ratio of the turbulent viscosity, $\nu_{T}=\alpha c_{g}^{2}/\Omega$, to the molecular viscosity of gas, $\nu_{m}=c_{g}\lambda/2$ (Cuzzi et al. 1993), with $\alpha$ the turbulence strength (Shakura \& Sunyaev 1973), $c_{g}$ the gas thermal speed, $\Omega$ the local Keplerian angular speed, and $\lambda$ the mean free path. $\mathrm{St}_i=\tau_i/t_L$ are the Stokes numbers of the two particles, with $\tau_i=\frac{3}{4c_{g}\rho _{g}}\frac{m_1}{\pi a_i^{2}}$ the stopping time of dust particle ($\rho_{g}$ is the gas density; $m_1$ and $a_1$ are the mass and equivalent radius of the particle), and $t_{L}=1/\Omega$ the turn-over time of the largest eddy. 
\\


For low-velocity collisions between $\mu$m-sized grains and (sub)mm-sized chondrules, i.e., $v<10$ cm s$^{-1}$, almost all collisions result in sticking at the point of contact (Ormel et al. 2008). With the introduction of charged particles, the electrostatic interaction exerts a force on the incoming particle. Charged particles will have reduced collision velocities due to the Coulomb repulsion, and particles with very low kinetic energies will not have enough energy to overcome the Coulomb barrier. Collisions with energies that exceed a certain minimum threshold will result in restructuring, bouncing, fragmentation or mass transfer (Wurm et al. 2005; Kothe et al. 2010), which pose other types of barriers to FGR growth. For simplicity, we limit the outcomes to sticking at the point of contact, restructuring, and repulsion (velocities are not large enough to cause bouncing in the simulation conditions used in this study; Dominik 1997).

\subsection{Charging of particles}\label{cha}

In many regions of a protoplanetary disk, the gas is weakly ionized and particles become charged due to the collection of electrons and ions on the surface.  The electron and ion currents to the grain surface are typically calculated using orbital-motion-limited (OML) theory (Allen 1992); the irregular aggregate surface is treated by modifying the OML currents using a line-of-sight approximation (OML-LOS) to adjust the currents to less-exposed regions of the surface (Matthews et al. 2012; Matthews et al. 2016). The rimmed chondrule is modeled as an aggregate of spheres. The surface of each sphere is divided into patches, and the charging current due to incoming plasma species is calculated for each patch. The current density of plasma species $\beta$ to a given surface patch $P$ depends on the number density $n_{\beta }$, charge $q_{\beta }$, and the velocity distribution $f(v_{\beta })$ of the plasma particles (Matthews et al., 2012),
\bqn
\lb{00}
J_{\beta,P } = n_{\beta }q_{\beta }\int \int \int f(v_{\beta })v_{\beta }cos(\theta )d^{3}\overrightarrow{v_{\beta }}
\eqn
where $v_{\beta }cos(\theta )$ is the velocity component of the incoming plasma particle perpendicular to the surface.

Since the lines of sight, i.e., paths for incoming electrons or ions, may be blocked by other monomers in the aggregate, the current density can be split into two parts: the component of the current density $J_{\beta }(v_{\beta })$, which only depends on the speed of the plasma particles, and the line of sight factor $L_{p}$ for a given patch,

\bqn
\lb{000}
J_{\beta,P } = J_{\beta0 } L_{p}.
\eqn
The integral over the velocities is given by
\lb{0000}
\bqn
J_{\beta0}=n_{\beta }q_{\beta }\int f(v_{\beta })v_{\beta }^{3}dv_{\beta }
=\left\{\begin{matrix}
\frac{n_{\beta }q_{\beta }}{\pi }\left ( \frac{k_{B}T_{\beta}}{2\pi m_{\beta}} \right )^{1/2} exp \left (- \frac{q_{\beta }\Phi _{P}}{k_{B}T_{\beta}} \right )~&\text{for}&~q_{\beta }\Phi_{P} \geqslant 0\\
\frac{n_{\beta }q_{\beta }}{\pi }\left ( \frac{k_{B}T_{\beta}}{2\pi m_{\beta}}\right )^{1/2}\left (1 - \frac{q_{\beta }\Phi _{P}}{k_{B}T_{\beta}} \right )~&\text{for}&~q_{\beta }\Phi _{P}< 0
\end{matrix}\right.
\eqn
\bqn
L_{p} = \int\int cos\left ( \theta  \right )d\Omega \lb{e23}.
\eqn
assuming a Maxwellian distribution for the ion and electron velocities. The minimum velocity a plasma particle must have to overcome the electrostatic repulsion and reach the dust particle surface is given by $v_{min} = \sqrt{2q_{\beta }\Phi _{P}/m_{\beta }}$ (with $\Phi _{P}$  the dust surface potential calculated from charge distributed over entire chondrule surface and $m_{\beta }$ the mass of the plasma particle) for $q_{\beta }\Phi _{P}>0$, and $v_{min}=0$ otherwise.  Integration over the angles gives the line of sight factor,

\bqn
\lb{0000}
L_{p} = \int\int cos\left ( \theta  \right )d\theta d\phi  \approx \sum_{t}LOS_{t,P}\times cos\theta_{t} \Delta \Omega
\eqn
The integral over the angles is approximated by a sum over test directions originating from the center of each patch $P$ which cover the solid angle $\Delta \Omega$. $LOS_{t,P}$ = 0, if a test direction is blocked by another monomer within the rim or the chondrule itself, and 1 otherwise. Monomers in the interior of the aggregate have small $L_{p}$, and thus collect little charge. The current reaching a given patch depends on the potential due to the charge of all other patches, and the charge on each patch is calculated in an iterative process until the equilibrium charge is reached (Matthews et al. 2012, 2016).

\begin{figure*}[!htb]
\includegraphics[width=7.5cm]{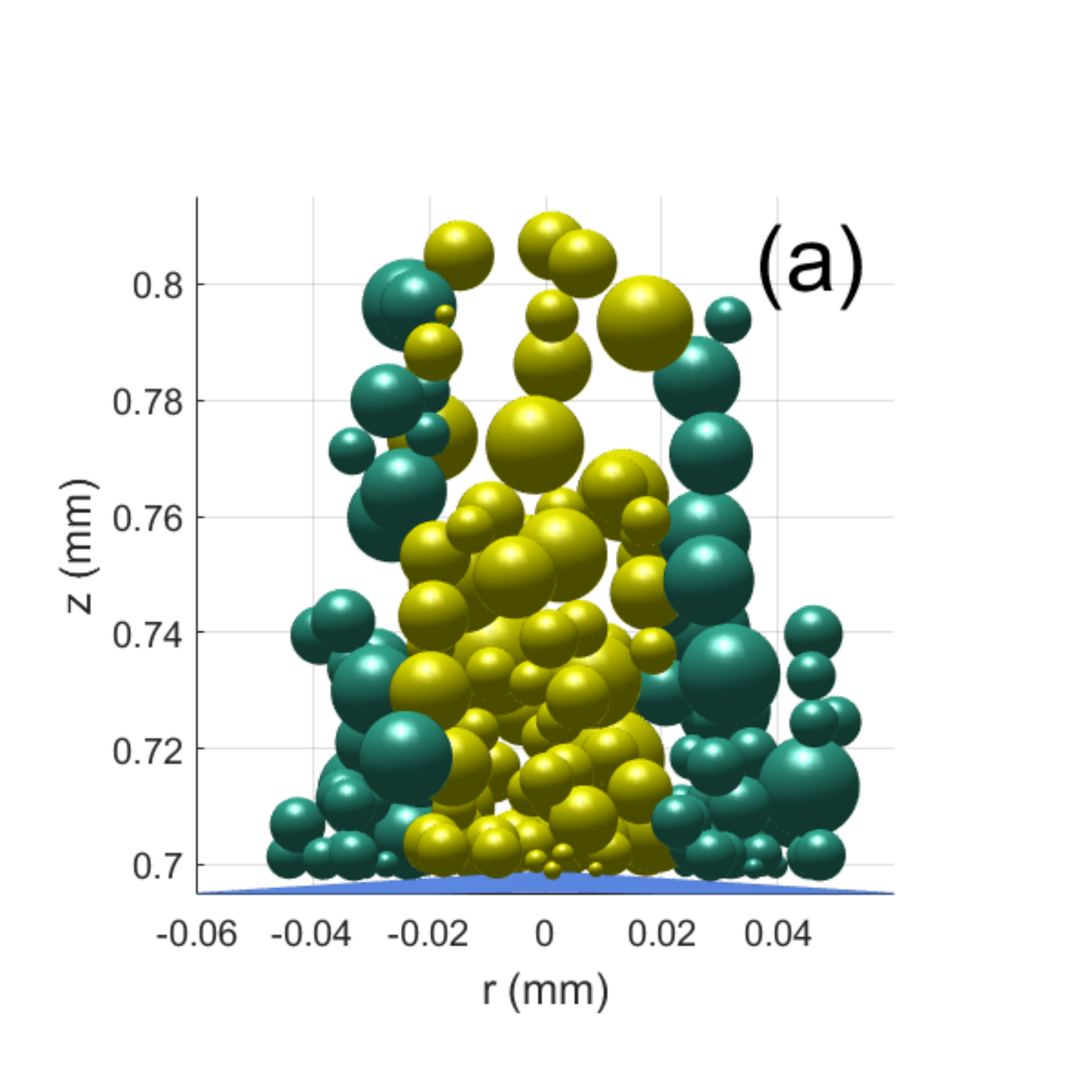}\includegraphics[width=7.5cm]{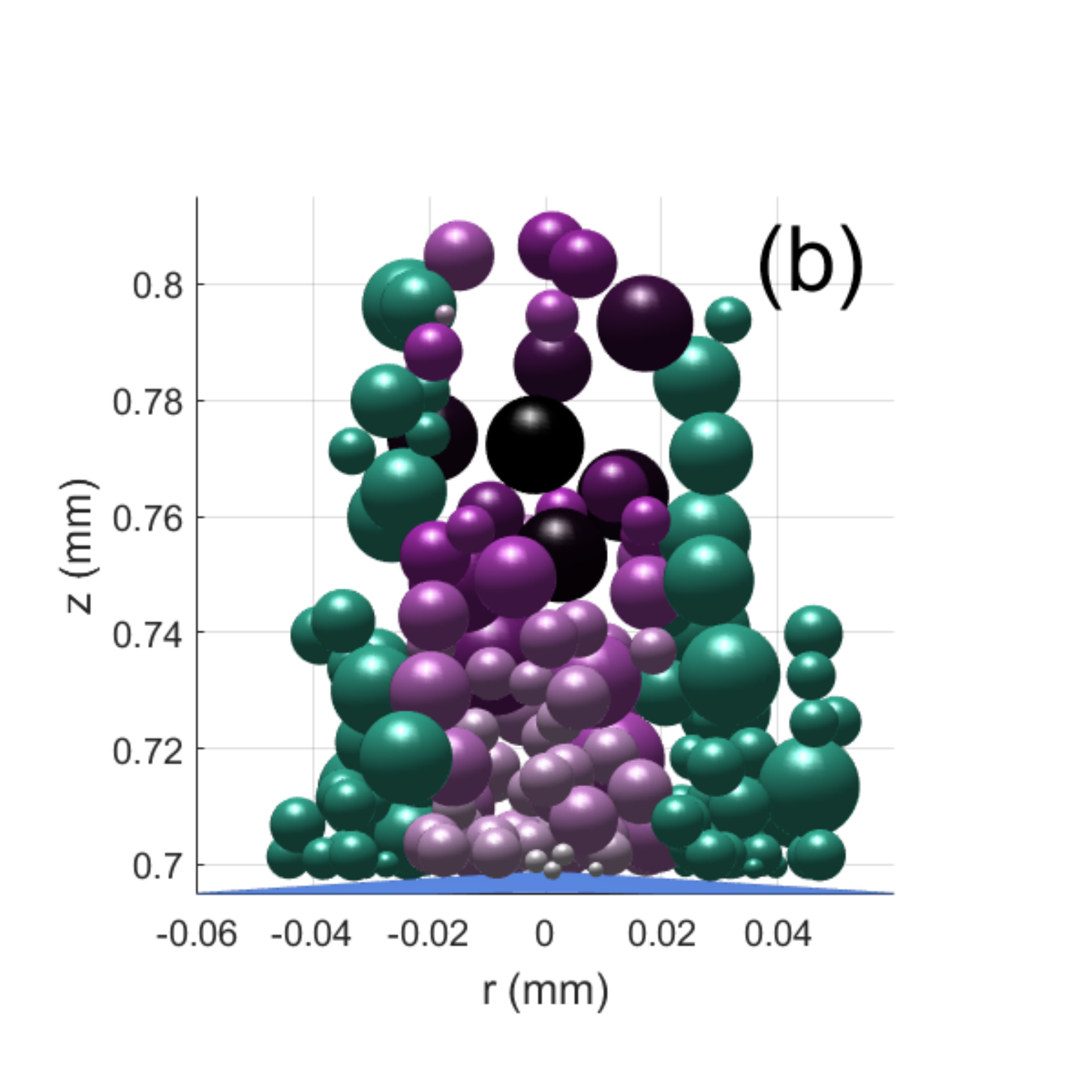} 
\includegraphics[width=7.5cm]{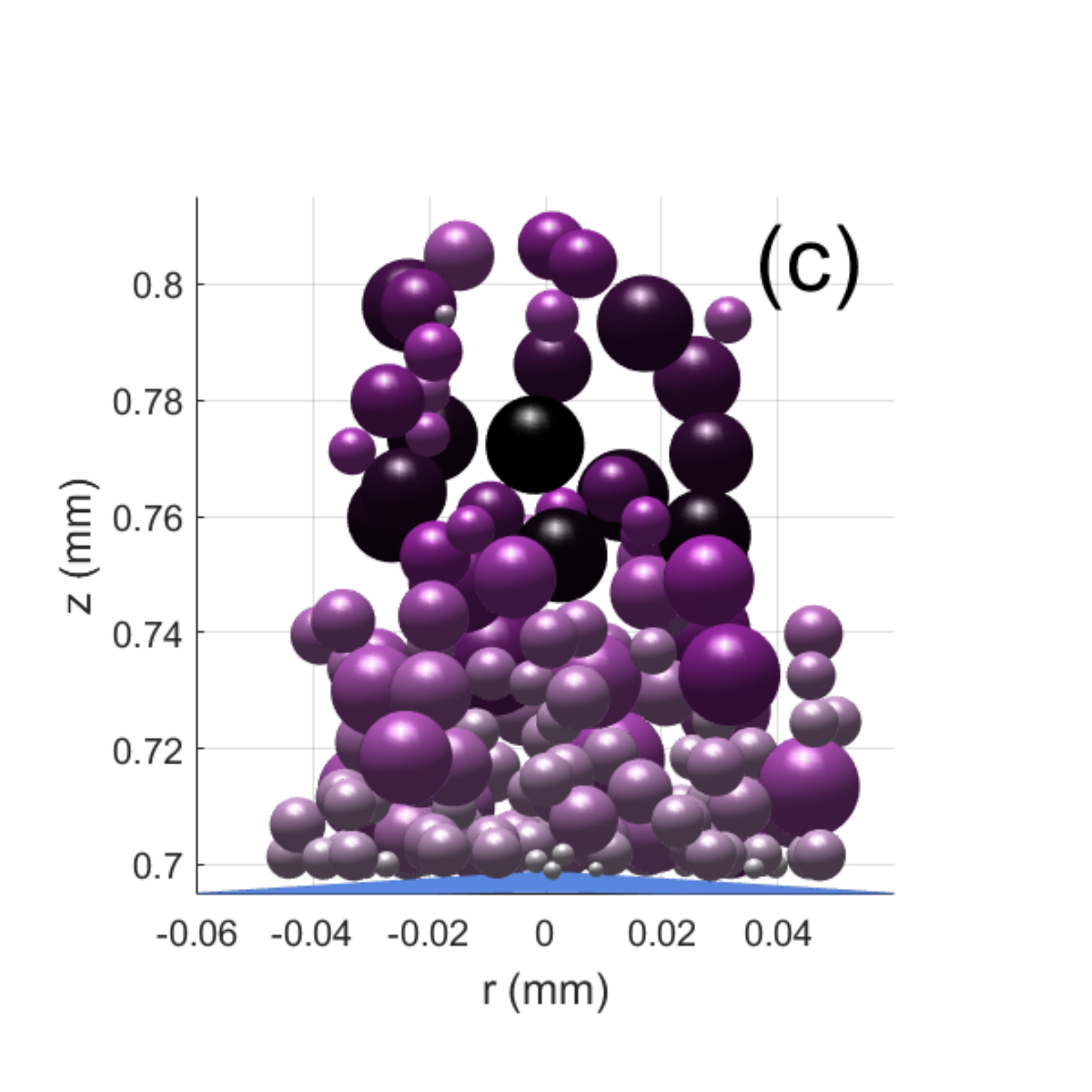}\includegraphics[width=7.5cm]{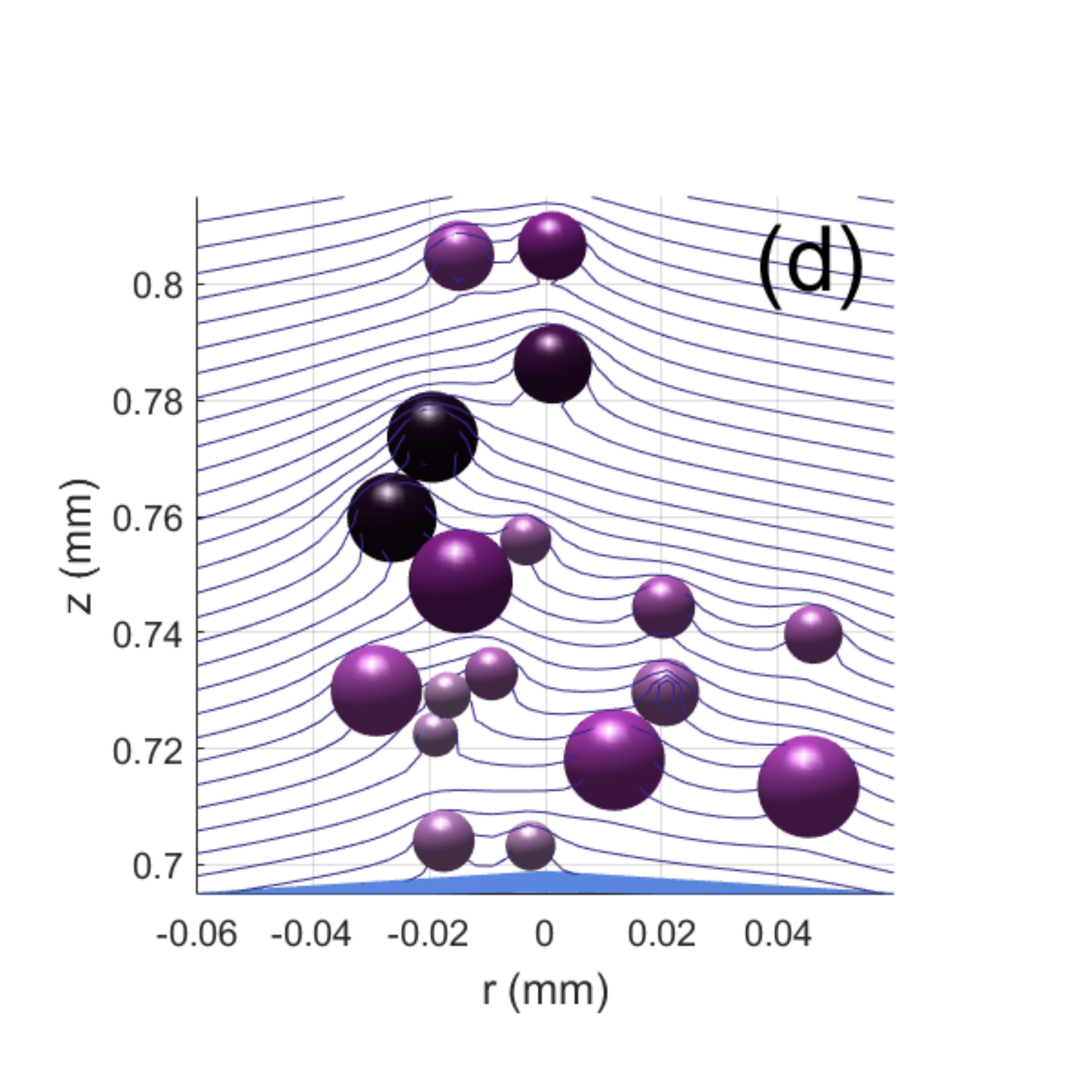}
\caption{Illustration showing two different regions used in the charging calculation and the resulting equipotential lines. a) depicts the initial state before the monomers are charged, with the two colors indicating the
inner (yellow) and outer (green) regions. b) represents the state after the monomers in the inner region have been charged. The charge collected by the inner spheres is calculated directly using OML\_LOS, and the outer spheres are used to calculate the LOS\_factor for the inner spheres. The resulting charge collected by each sphere in the inner region is indicated by the shade: large spheres near the top of the pile collect the most charge (dark), while small, interior spheres collect little charge (light).  c) represents the state after all monomers have been charged. The charges on the inner monomers are used to approximate the charge on the spheres in the outer region, based on their size and distance from the chondrule surface. d) shows equipotential lines for a slice through the center of the dust pile.}
\label{pot}
\end{figure*}

The charge on the rimmed chondrule includes two parts: the charge on the spherical chondrule core and the charge on the porous dust rim. As we only simulate a dust pile on the chondrule, the monomers on the vertical sides of the pile have a greater $L_{p}$ than if the entire dust rim were included in the simulation. Therefore, we divide the dust pile into two regions, an inner region (the yellow spheres in figure \ref{pot}a) and an outer region (the green spheres in figure \ref{pot}a), and only use OML\_LOS to find the new charge on the monomers in the inner region. The outer monomers are included in the calculation of $L_{p}$ for the inner monomers (figure \ref{pot}b). After the charges on the monomers in the inner region are obtained, they are used to approximate the charges on the outer monomers: the monomers in the two regions are sorted separately based on their distance $z$ from the top of the dust pile, and the charges on the inner monomers are assigned to the outer monomers with the same ranking, scaled by the ratio of the radius of the outer monomer to that of the inner monomer (figure \ref{pot}c),

\bqn
\lb{0000}
Q_{out}(z)=Q_{in}(z)\times \frac{r_{out}}{r_{in}}
\eqn
where $Q$ and $r$ are the charge and radius of a monomer, with the subscripts ``out'' and ``in'' representing outer and inner monomers, respectively. In this manner, the charge on the monomer is related to its radius and distance from the chondrule surface. After the charges on the monomers are obtained, the total charge of the dust pile and the chondrule patch beneath the pile are duplicated on the rest of the chondrule surface, based on the ratio of the chondrule patch area to the total chondrule surface area. The charge on the entire chondrule (used to determine the trajectory of dust particles) is then calculated by summing the charge on each section of the chondrule. As shown in Fig. \ref{pot}d, the deviation from spherical equipotential lines is very small except very near to the dust rim surface.  
\\

At large distances, the electrostatic force acting on the incoming dust is calculated using a multipole approximation up to the quadrupole terms, where the multipole moments are calculated from the charge on the chondrule core and the distribution of charge on the dust rim.  At close approach, the charge approximation using the dipole and quadrupole moments breaks down, and the electrostatic force on the incoming dust is calculated using the total charge on each spherical monomer plus the charge on chondrule surface, treating the monomers as point charges [see Matthews et al. 2016].

\subsection{Physical parameters used in the simulation}

Dust coagulation was modeled for conditions at the midplane of the minimum-mass solar nebula (Hayashi 1981) at a distance of 1 AU from the sun, with a temperature of 280 K. The average molar mass, 2.33 g/mol, sound speed, $1.179\times 10^{5}$ cm/s and molecular viscosity of the gas, $1.8\times 10^{-4}$ g/(cm$\cdot$s), are used to calculate the relative velocity between the dust particles and chondrule by Eq. \ref{3}.


There are several mechanisms that can contribute to the ionization of the protoplanetary disk, such as cosmic rays, UV radiation, X-rays and the decay of radionuclides. In our simulation, the plasma environment is assumed to be hydrogen with equal electron and ion temperature, $T_{e}$ = $T_{i}$= 280 K, as the plasma thermalizes with the gas due to collisions. In the case of low dust density, a negligible percentage of the electrons reside on the dust grains, and the number density of electrons and ions in the gas is set to be $n_{e}=n_{i}=3.5\times 10^{8} \ m^{-3}$ (Horanyi \& Goertz, 1990). For high dust density, the ratio of free electrons to free ions is reduced due to electron depletion, which is modeled by varying the ratio $n_{e}/n_{i}$. The corresponding surface potentials of spherical grains are -0.061 V, -0.048 V and -0.020 V for $n_{e}/n_{i}=$ 1, 0.5, and 0.1 respectively. We compare rims formed in these three plasma environments to those formed in a neutral environment where the particles are not charged.

The precise value of the turbulence strength in protoplanetary disks, as quantified by the $\alpha$ parameter, is uncertain, and the values often considered range from $\sim 10^{-6}$ to 0.1 (Hartmann et al. 1998; Cuzzi 2004; Ormel et al. 2008, Carballido 2011). In this study, we investigate levels of turbulence with $\alpha$ equal to $10^{-1}, 10^{-2}, 10^{-3} , 10^{-4}, 10^{-5}, 10^{-6} $.

Chondrule diameters range from $\sim 100\ \mu m$ to $\sim 100\ \mu m$, depending on chondrite group, with approximate log-normal distributions (Friedrich et al. 2015). Here we simulate chondrules with radii between 500 and 1000 $\mu$m, in 100-$\mu$m increments. The radii of dust particles are taken to be $0.5 \sim 10\ \mu m$ with a power law distribution $n(r)dr\propto r^{-3.5}dr$. 


\section{RESULTS}

\renewcommand{\theequation}{3.\arabic{equation}} \setcounter{equation}{0}

\begin{figure*}[!htb]
\includegraphics[width=9cm]{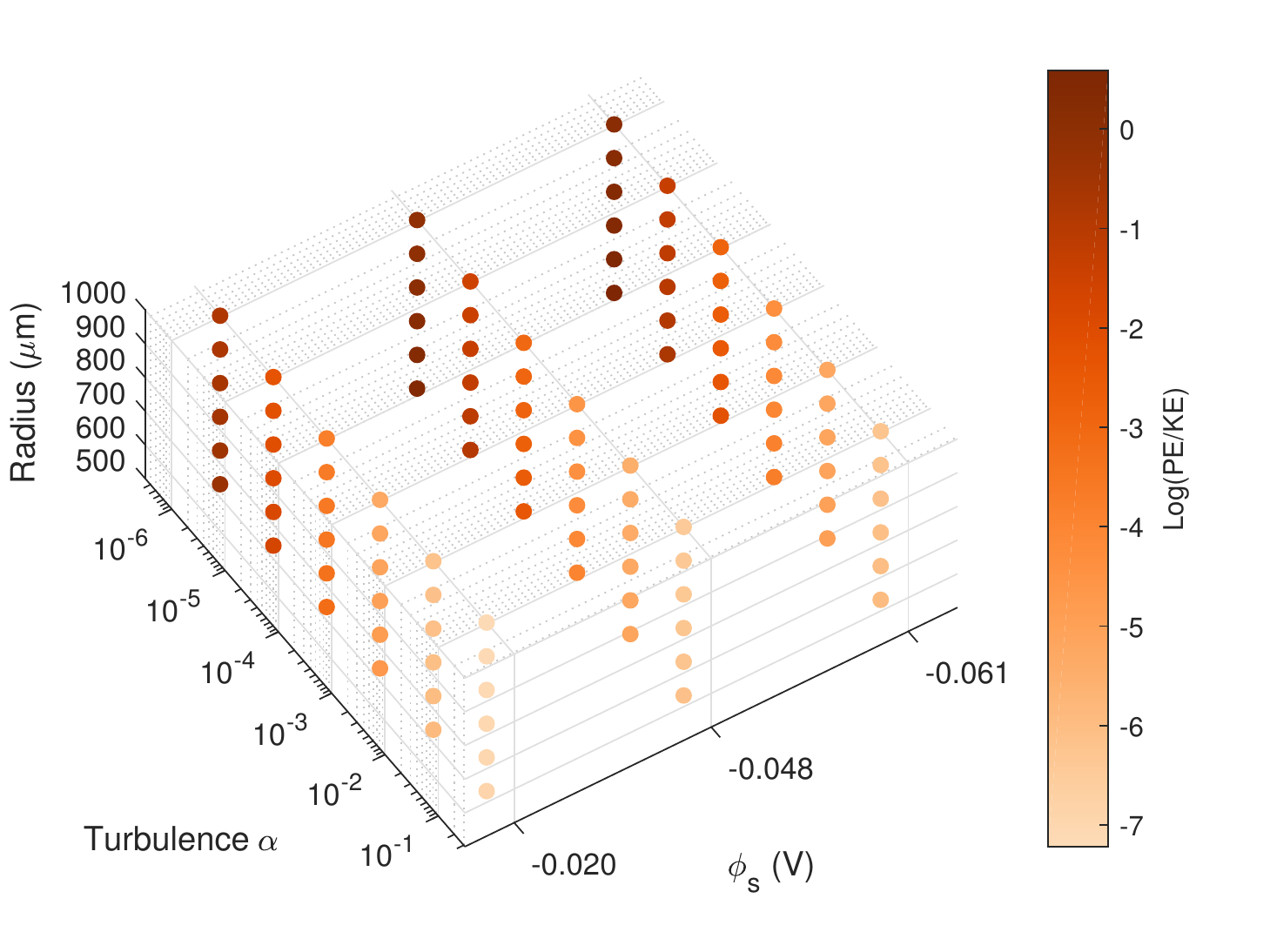} 
\caption{PE/KE (the ratio of a grain's electrostatic potential energy upon collision to its kinetic energy at large distances), as a function of dust surface potential ($\phi _{s}$ = -0.061 V, -0.048V, -0.020 V), turbulence strength ($\alpha=10^{-1}$, $10^{-2}$, $10^{-3}$, $10^{-4}$, $10^{-5}$, $10^{-6}$) and chondrule size ($r$ = 500-1000 $\mu$m, in 100-$\mu$m increments).}
\label{f709}
\end{figure*}

The structure of the dust rims and their formation time are affected by the level of turbulence, charging condition and the chondrule size. Together, these effects can be characterized by the ratio PE/KE, where $PE$ is the electrostatic potential energy of the dust grain at the point of impact, given by $PE=\frac{\epsilon_{0}Q_{d}Q_{c}}{4r_{dc}}$, with $Q_{d}$ and $Q_{c}$ the charges of the dust grain and the chondrule and $r_{dc}$ the distance between the dust grain and the center of the chondrule, and the kinetic energy of the dust grain far from the chondrule is $K_{E}=\frac{1}{2}m_{d}v_{rel}^{2}$, with $m_{d}$ the mass of the dust grain and $v_{rel}$ the relative velocity calculated from Eq. \ref{3}. Figure \ref{f709} shows the range of PE/KE averaged over the dust population for different chondrule sizes in different environments (turbulence strength; plasma conditions). Given the combined turbulence and charge levels, the results are broadly applicable over a large region of the disk where conditions match a given PE/KE. The structure of the resulting dust rims is characterized based on the distribution of monomer sizes, porosity ($\psi $), and the time for rim formation as a function of PE/KE.

\subsection{Size distribution of dust collected in the rim}

\begin{figure*}[!htb]
\includegraphics[width=18cm]{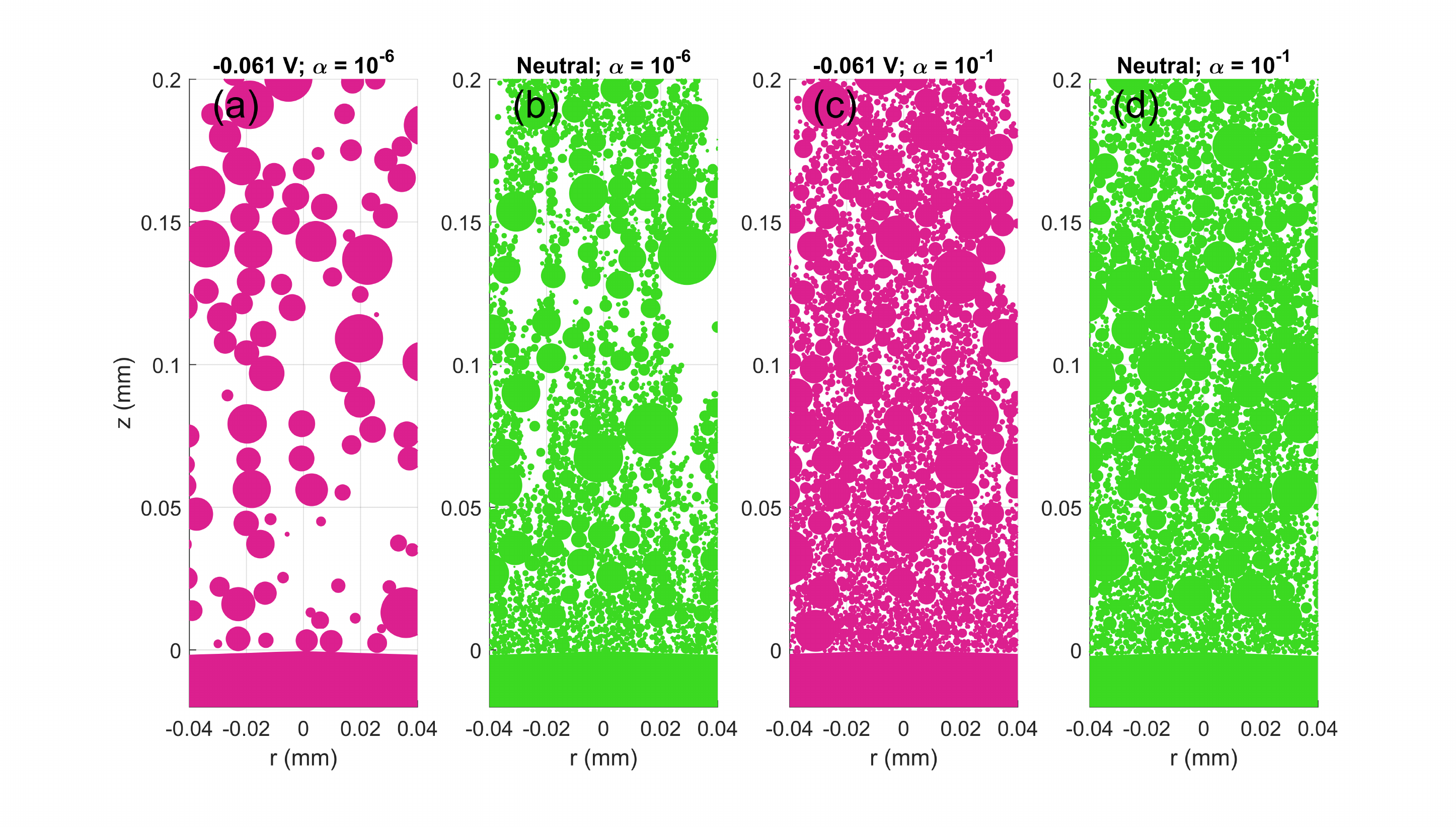}
\caption{Cutaway view of rims collected on a 100-$\mu$m-diameter patch on the surface of a chondrule with a radius of 1000 $\mu$m, formed from a) charged dust ($\phi _{s}$ = -0.061V) in weak turbulence $\alpha=10^{-6}$; b) neutral dust in weak turbulence $\alpha=10^{-6}$; c) charged dust ($\phi _{s}$ = -0.061V) in strong turbulence $\alpha=10^{-1}$; d) neutral dust in strong turbulence $\alpha=10^{-1}$. The size of each circle represent the apparent monomer size due to cutting effects. }
\label{f1}
\end{figure*}

Although the population of dust in the protoplanetary disk has a power law size distribution with average grain radius $a_0 \approx 0.83\ \mu m$, the electrostatic repulsion alters the distribution of grains collected within the chondrule rim. The distribution of monomer size within the rim depends not only on the magnitude of the charge, but also on the relative velocities between the dust and chondrule.  Representative slices through the dust rim are shown in Fig. \ref{f1} comparing the results for two levels of charge and turbulence. In a weakly turbulent environment, where the relative velocities are low, only the largest dust grains have enough energy to overcome the Coulomb repulsion barrier (Fig. \ref{f1}a), whereas for the same turbulence level, a rim consisting of uncharged grains will consist of dust particles of all sizes (Fig. \ref{f1}b), and the smaller dust grains are able to penetrate the porous dust pile and are concentrated near the chondrule surface. In a highly turbulent region where the relative velocities are large, the distribution of dust grain sizes is similar for both charged and neutral grains (Fig \ref{f1}c, \ref{f1}d). However, as shown below, the distribution of particle sizes within the rim and the porosity of the rims differ.

\begin{figure*}[!htb]
\includegraphics[width=9cm]{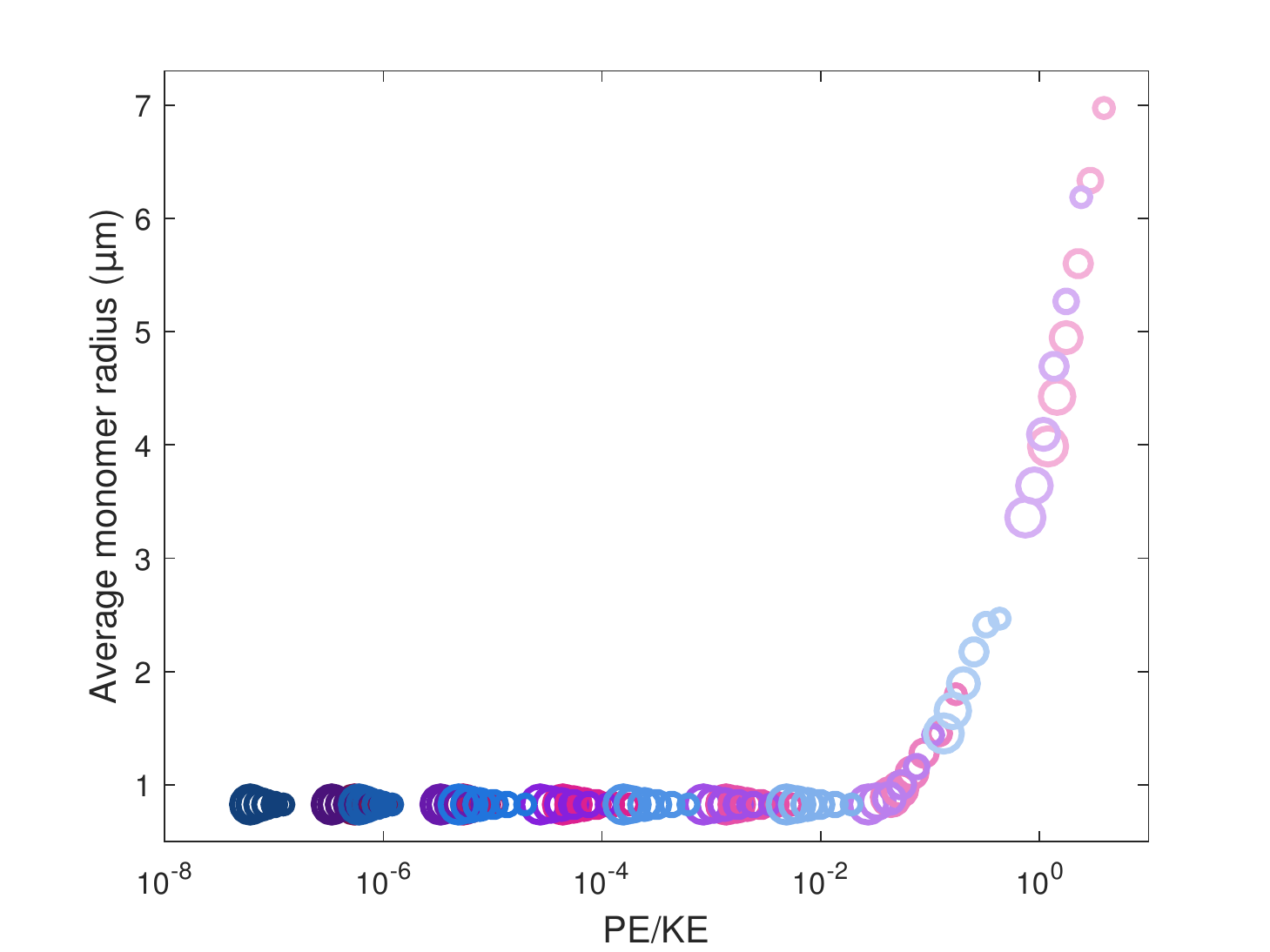} 
\caption{Comparison of average dust size within dust rims with a thickness of 200 $\mu$m, as a function of PE/KE. Dust surface potential is indicated by color (pink: $\phi _{s}$ = -0.061V; purple: $\phi _{s}$ = -0.048V; blue: $\phi _{s}$ = -0.020V). Turbulence level is denoted by shade($\alpha=10^{-1}$ to $\alpha=10^{-6}$ in order of decreasing color shades). Size of chondrule core is represented by symbol size (r = 500-1000 $\mu$m, in 100 $\mu$m increments).}
\label{f7}
\end{figure*}

The average monomer size within dust rims when the rims have reached a thickness of 200 $\mu$m in environments with different turbulence levels ($\alpha=10^{-1} - 10^{-6}$) and charge levels ($\phi _{s}$ = -0.061V, -0.048V, -0.020V) are compared in Fig. \ref{f7}. For low values of PE/KE, the presence of charged grains has little effect, and the average monomer size within the dust rims matches that of the original dust population. Above a critical value of PK/KE $\approx 0.0193$, the average monomer size within the rims increases with PE/KE, as the smallest dust particles are repelled from the chondrules. For a given turbulence level (indicated by shade), and charge level (indicated by color), small chondrules (indicated by symbol size) tend to collect larger monomers. 


\begin{figure*}[!htb]
\includegraphics[width=9cm]{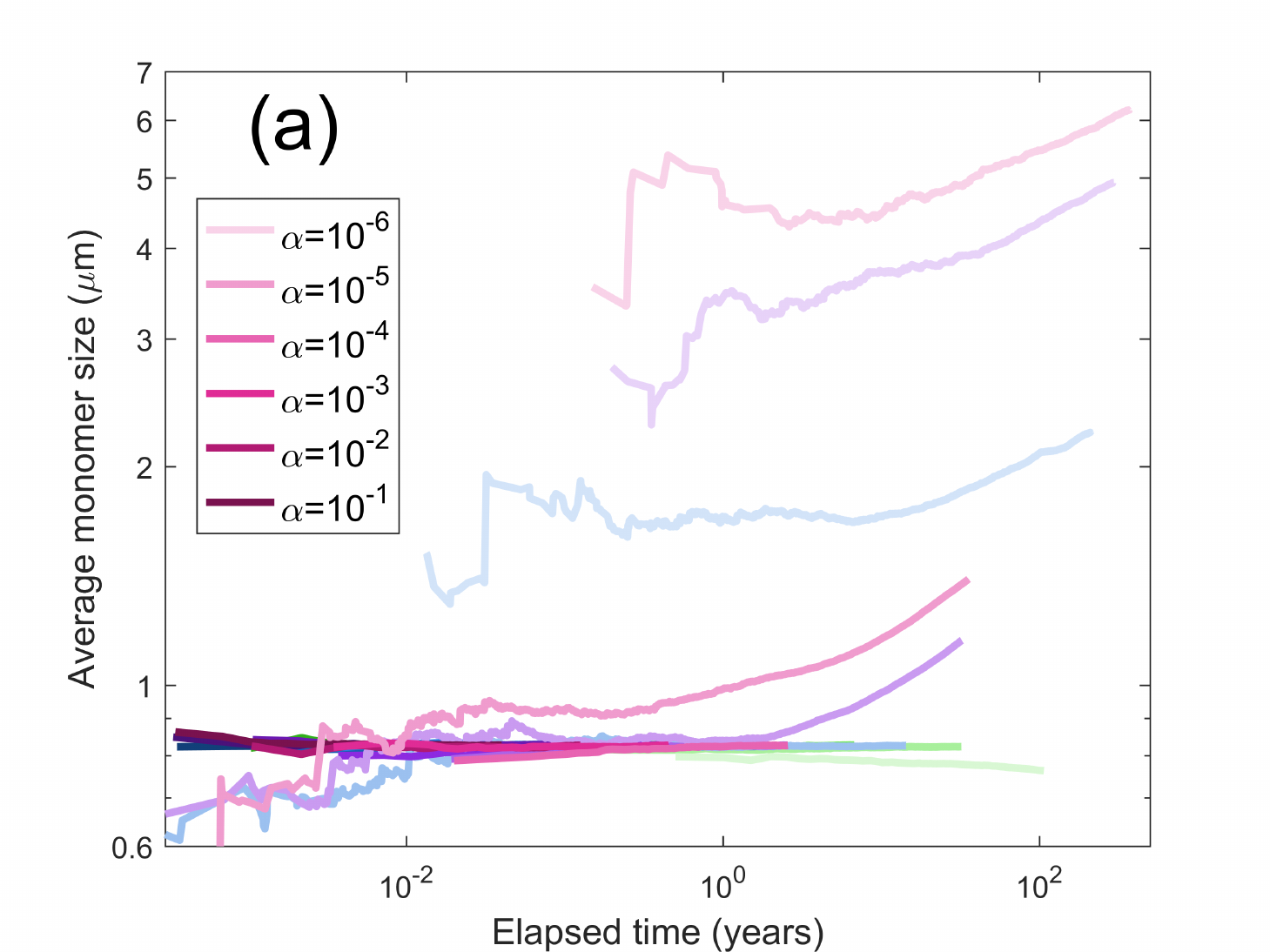}\includegraphics[width=9cm]{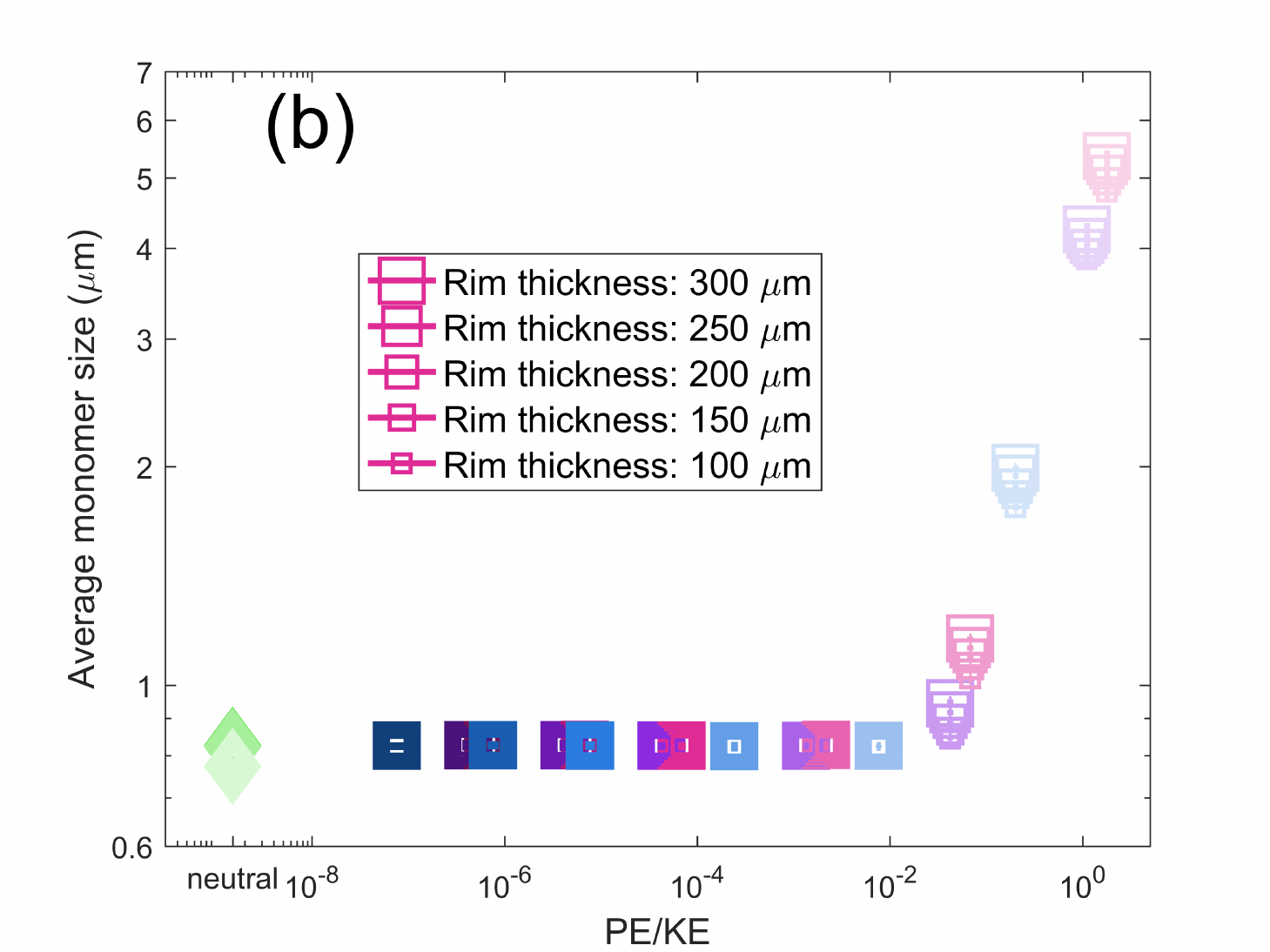}  
\caption{Average monomer size within dust rim on a 900-$\mu$m-radius chondrule, a) as a function of elapsed time, and b) as a function of PE/KE. Data is shown for all turbulence levels ($\alpha=10^{-1}$ to $\alpha=10^{-6}$ in order of decreasing color shades) and dust surface potentials (pink: $\phi _{s}$ = -0.061V; purple: $\phi _{s}$ = -0.048V; blue: $\phi _{s}$ = -0.020V). In (b), the symbol size indicates the total thickness of the rim as the rim grows from 100 $\mu$m (small squares) to 300 $\mu$m (big squares). For comparison, the results for neutral dust particles are shown in green.}
\label{f8}
\end{figure*}

The time evolution of the average dust size collected as the rim grows in thickness is shown in Fig. \ref{f8}a for a representative chondrule with a radius of 900 $\mu$m. Neutral chondrules collect dust with the average size matching that of the population for all turbulence levels. In charged environments, the average size of the dust contained in the rims remains nearly constant for turbulence $\alpha\geq 10^{-4}$ but increases with rim growth for lower turbulence levels. This indicates that the repulsion of smaller dust grains increases as the rimmed chondrule grows and collects more charge. The average monomer size as the rim grows is also shown as a function of PE/KE in Fig. \ref{f8}b. Again, as in Fig. \ref{f7}, PE/KE $\approx 10^{-2}$ is a critical threshold above which the average monomer size increases with PE/KE. As the rim grows thicker, as indicated by the symbol size, more large dust grains are incorporated.


\begin{figure*}[!htb]
\includegraphics[width=6cm]{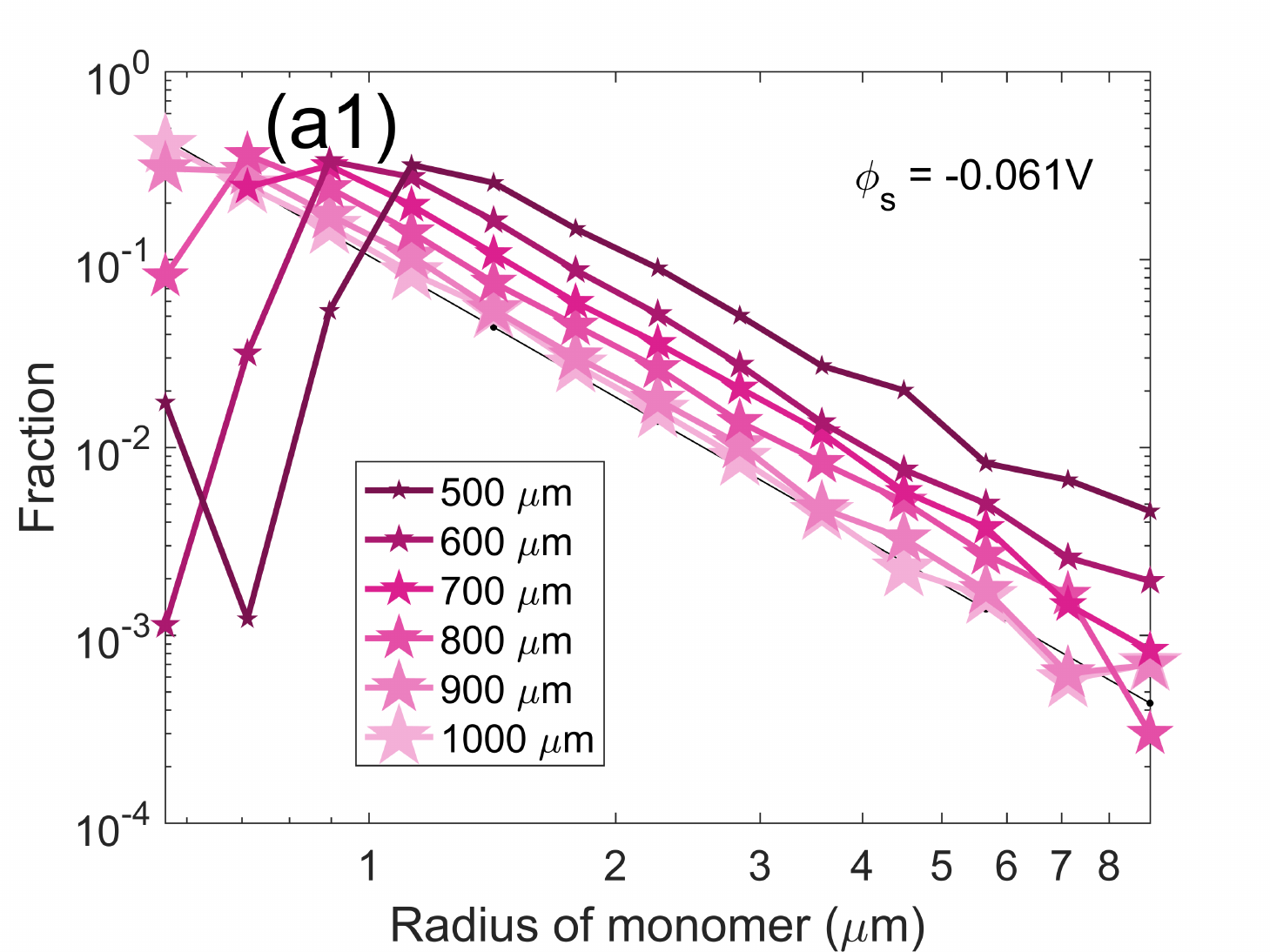}\includegraphics[width=6cm]{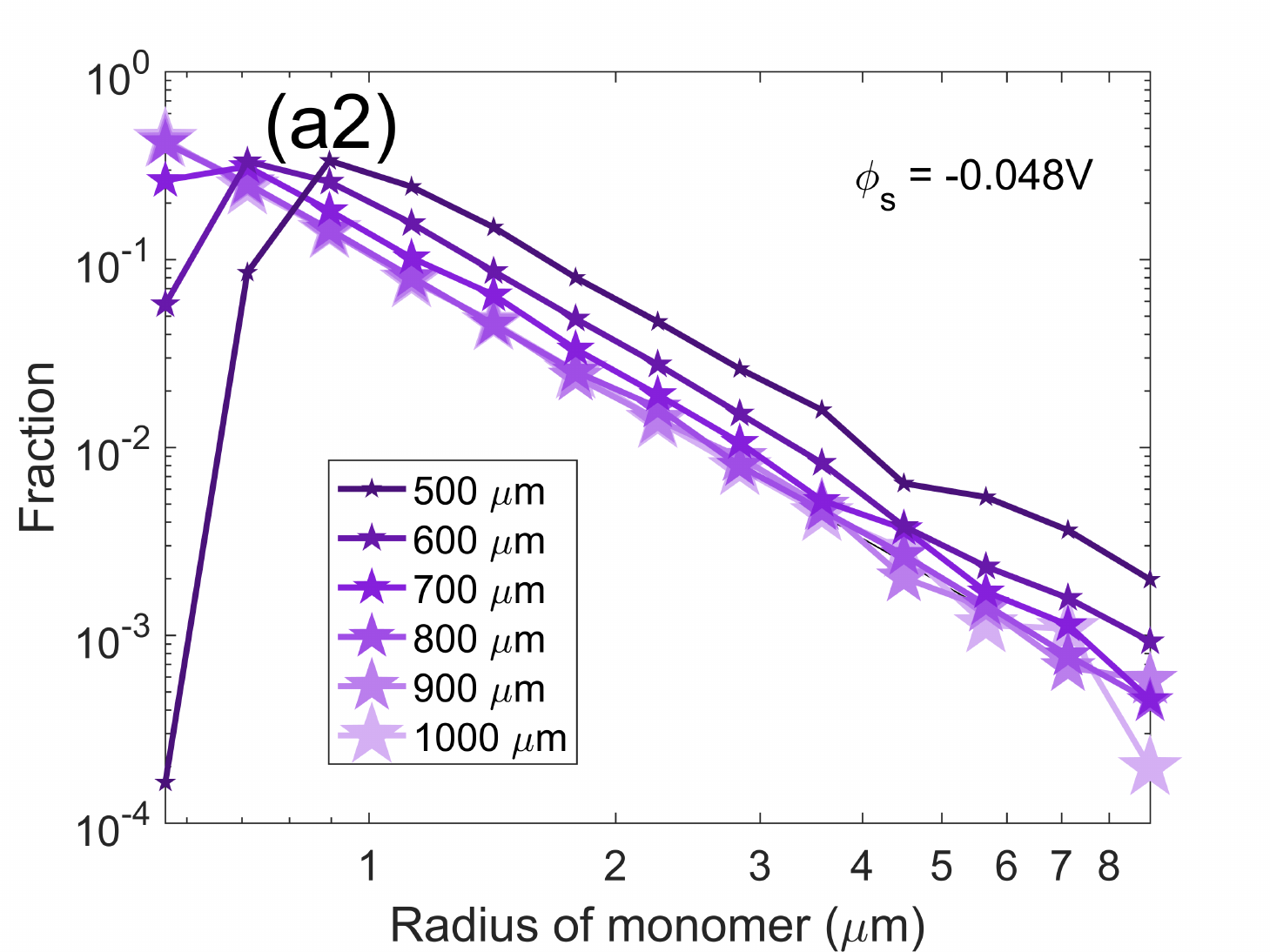}\includegraphics[width=6cm]{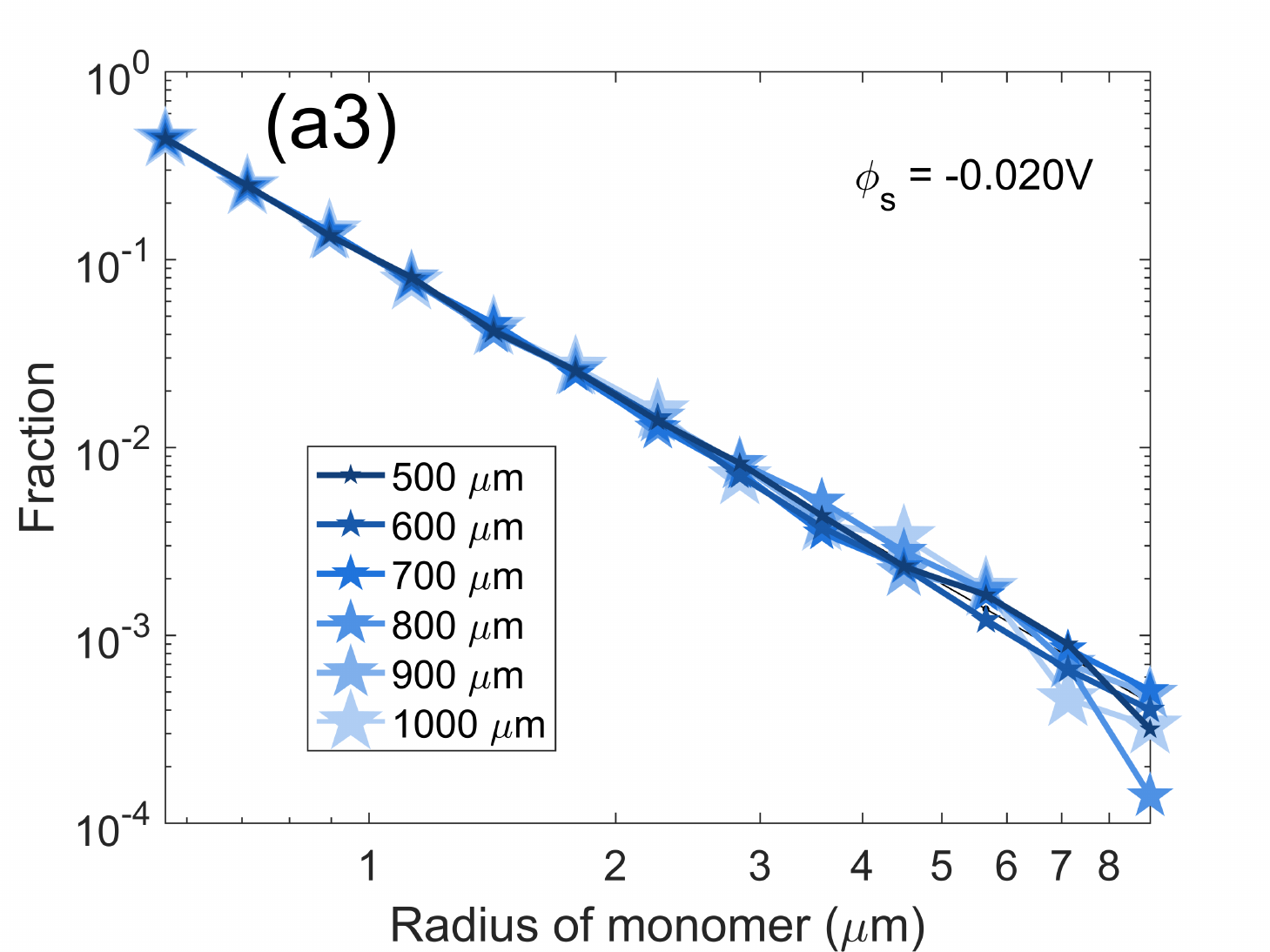} 
\includegraphics[width=6cm]{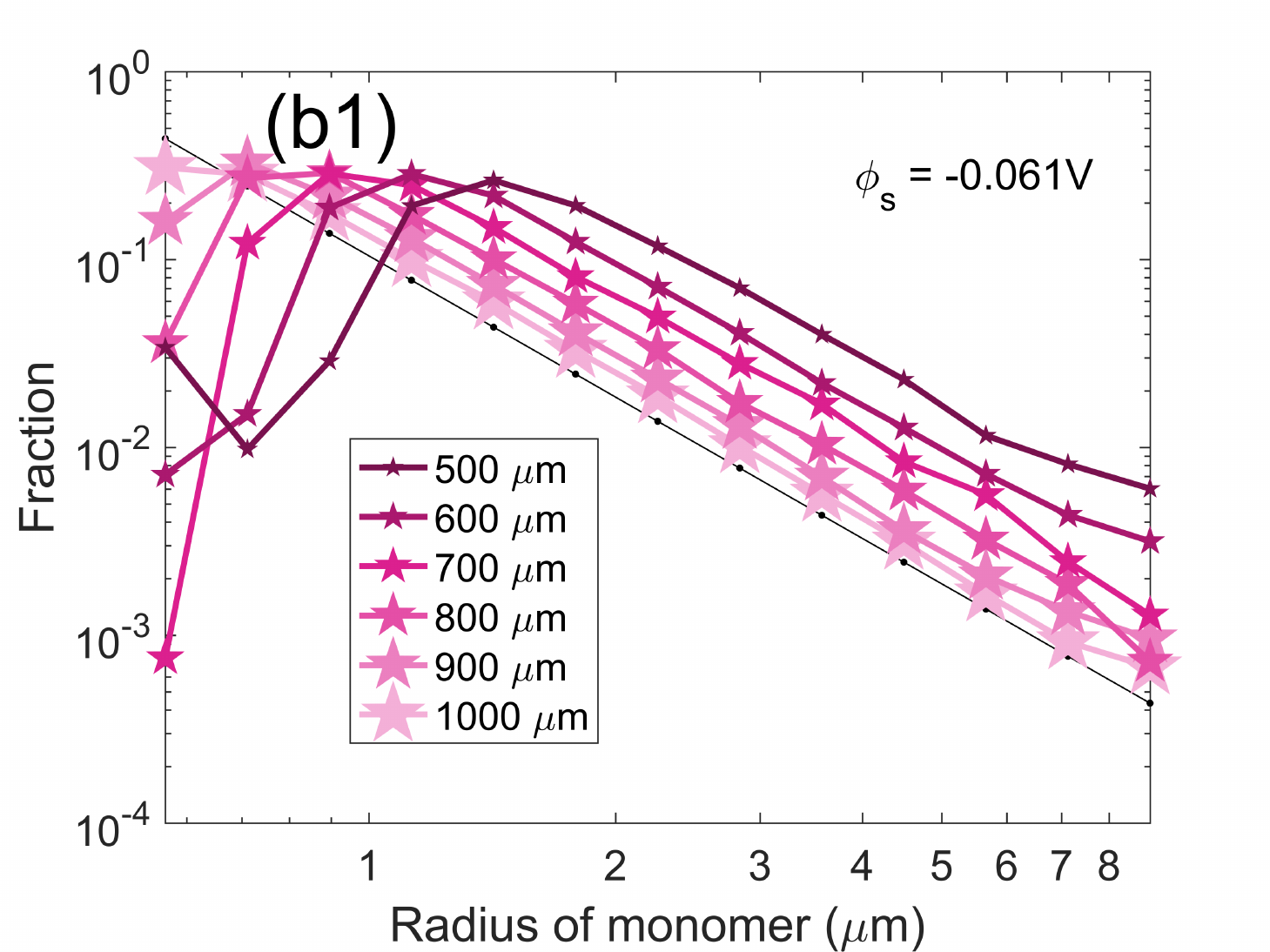}\includegraphics[width=6cm]{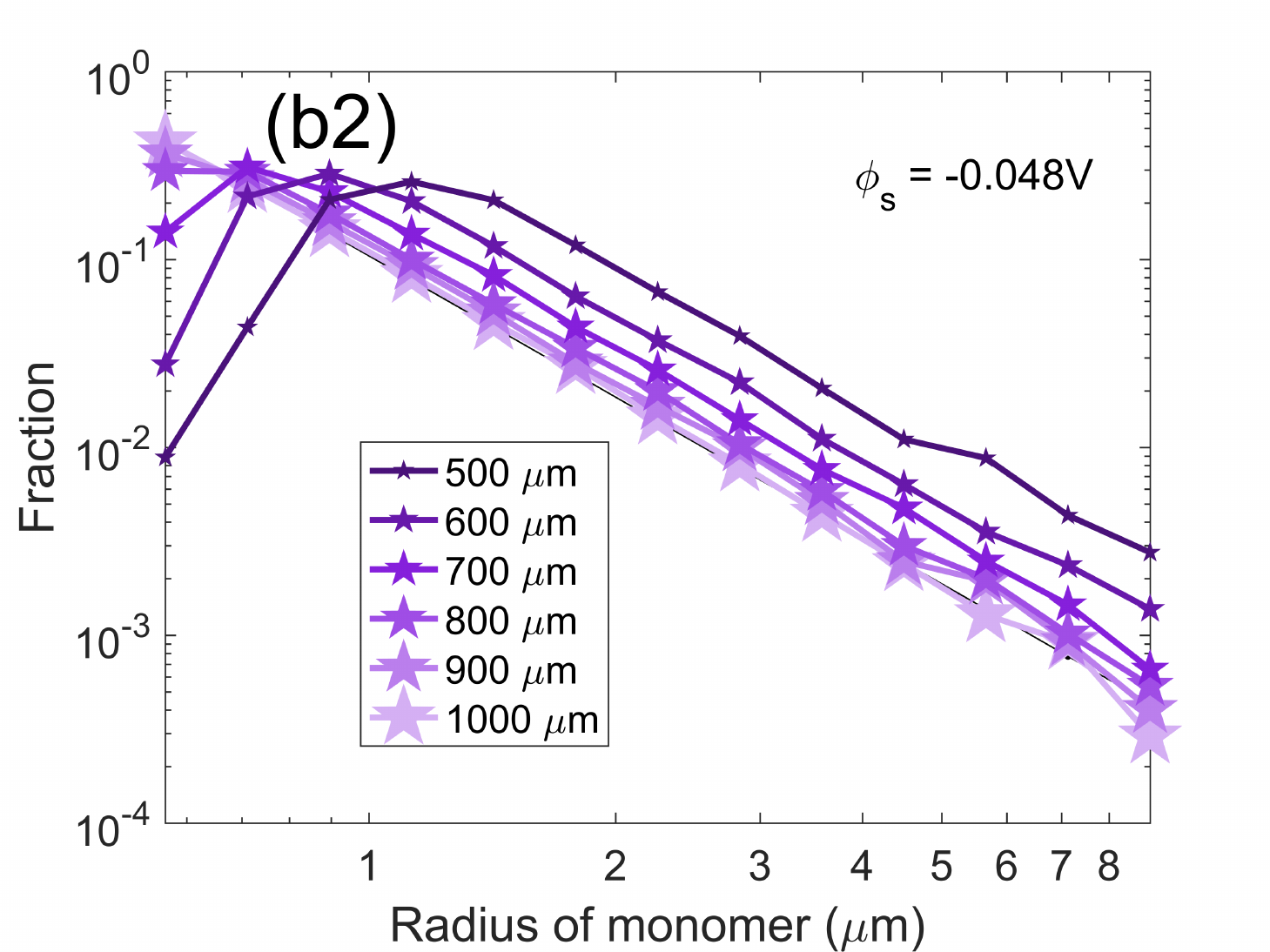}\includegraphics[width=6cm]{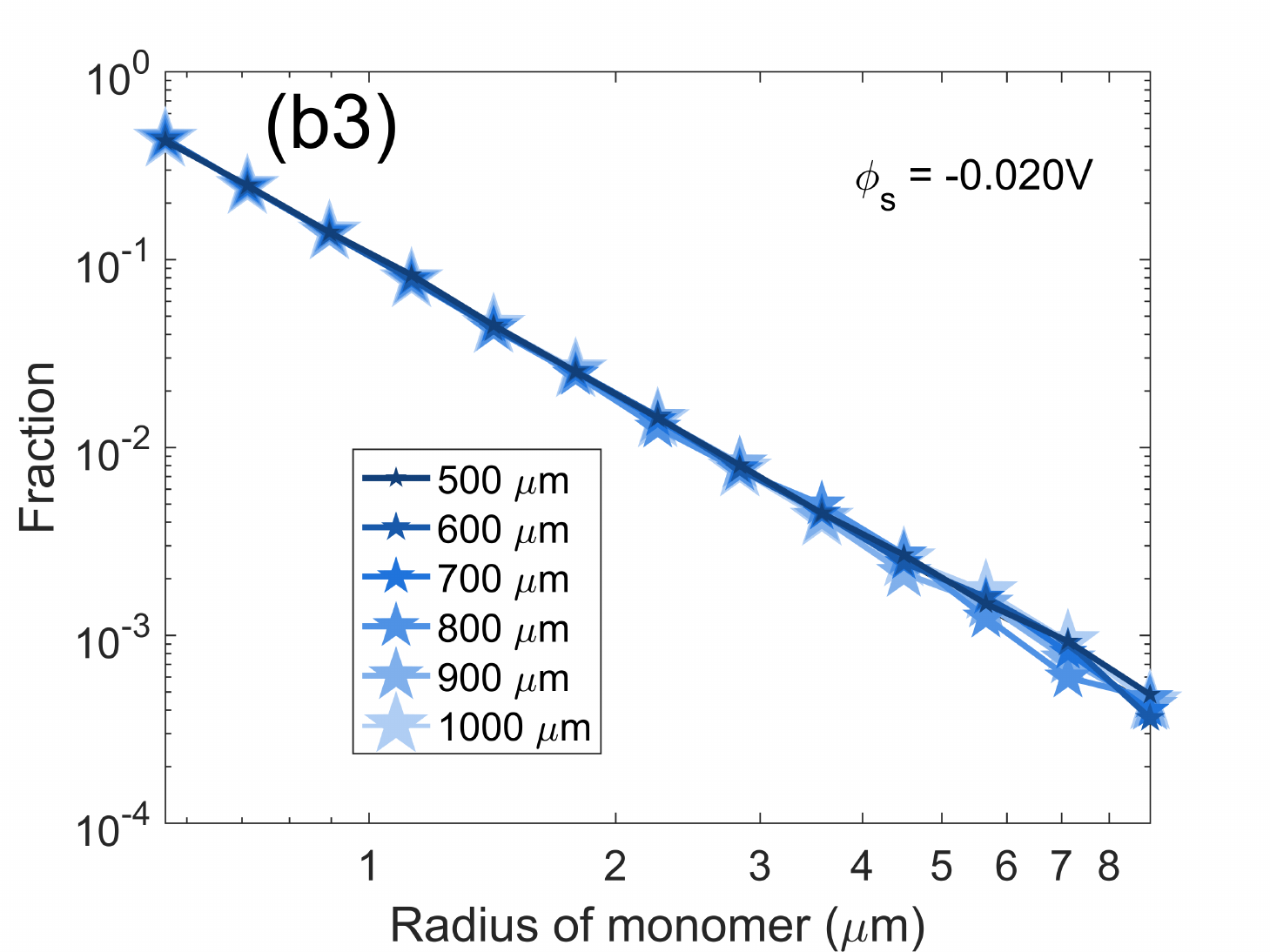}
\includegraphics[width=6cm]{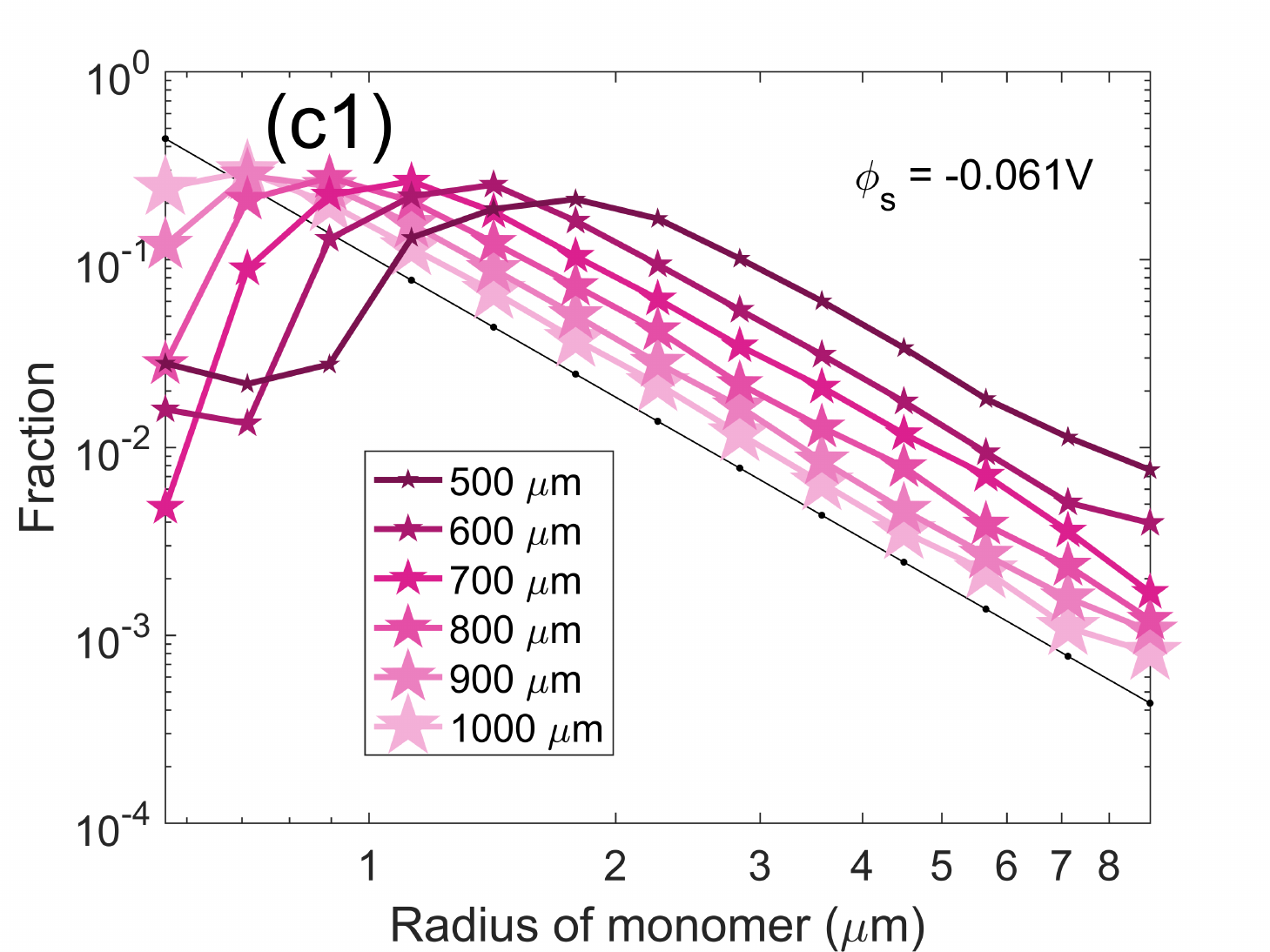}\includegraphics[width=6cm]{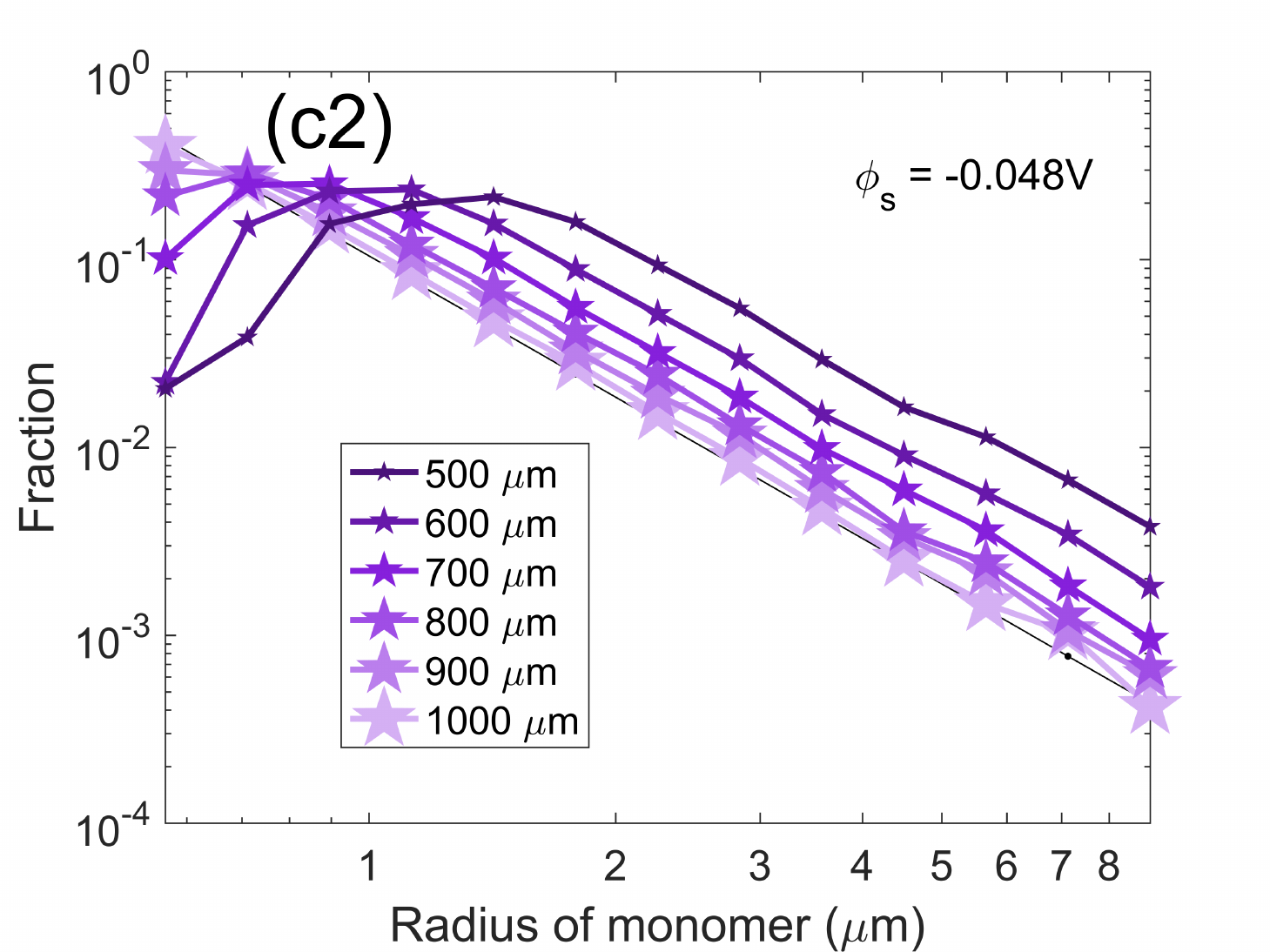}\includegraphics[width=6cm]{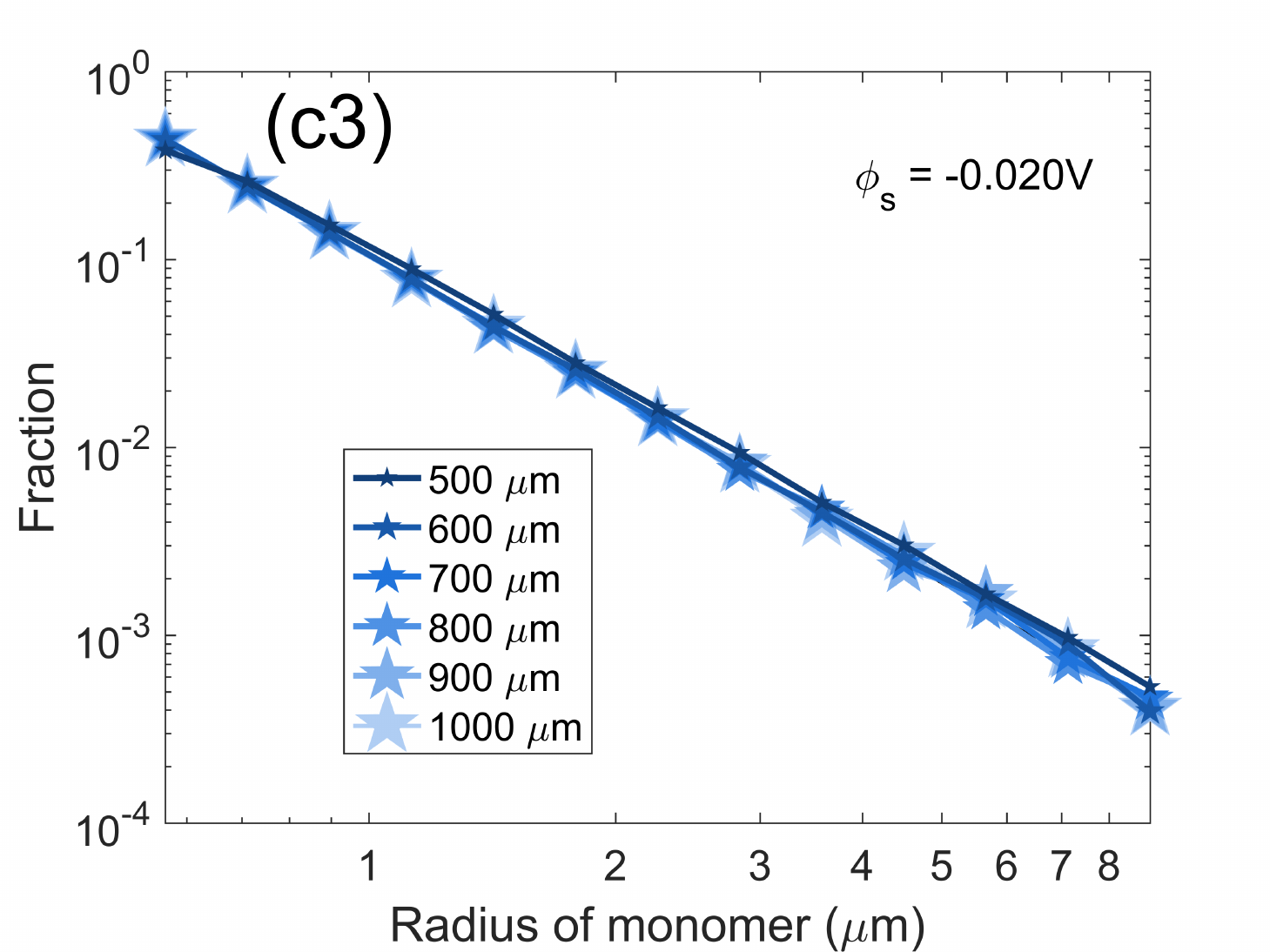}
\caption{Monomer size distribution in the dust rims as the rim thickness grows from a) 100 $\mu$m, to b) 250 $\mu$m to c) 400 $\mu$m.  The dust surface potential is (column 1, pink) $\phi _{s}$ = -0.61 V, (column 2, purple) $\phi _{s}$ = -0.048 V, and (column 3, blue) $\phi _{s}$ = -0.020 V.  The initial chondrule radius is indicated by the symbol size, as given in the legend. The turbulence strength is $\alpha=10^{-4}$. The distribution of the initial dust population is shown by the thin black line.}
\label{f9}
\end{figure*}

To further illustrate the difference that charge plays in weak turbulence, the distribution of monomer sizes for the three different levels of surface potential at three different stages in the rim growth is presented in Fig. \ref{f9} for turbulence level $\alpha=10^{-4}$. The distribution shifts towards larger monomers as a dust rim grows in thickness in the highly charged environments ($\phi _{s}$ = -0.061 V and -0.048 V), while the change is minor in the environment with a low dust surface potential ($\phi _{s}$ = -0.020 V). Overall, large monomers with radii greater than 1 $\mu$m are more common in small chondrules, in environments with high surface potentials, and in thicker rims (those that have more time for accumulation).

\subsection{Porosity}
The porosity is a measure of the open space within the dust rim. It can be examined in experimental observations of meteorite samples (Greshake et al. 2005; Friedrich 2014), and, along with other features of the fabric, provide important information about the environmental conditions where the dust rims were originally formed and the subsequent processes that alter the porosity (Thompson 1985; Gunkelmann et al. 2017). Differences can be seen not only in the average porosity of the entire rim, but also in the change in the porosity of different layers within the rim. To calculate the porosity as a function of the distance from chondrule surface, the rim is divided into a certain number of horizontal layers (i.e., parallel to the chondrule’s surface), with each layer having a thickness of 3--4 $\mu$m. The porosity of each layer is defined as the ratio of the volume of voids within the layer (the total volume of the layer $V_{layer}$ minus the sum of the volume of the monomers (or monomer portions) within that layer $\sum V_{d}$) to the total volume of the layer, i.e., $\left (V_{layer}-\sum V_{d}  \right )/V_{layer}$. To avoid edge effects of the dust pile, only the inner region of the pile with a radius of 50$\%$ of the total pile radius (i.e., half the distance from the pile center to pile edge) is analyzed.

\begin{figure*}[!htb]
\includegraphics[width=6cm]{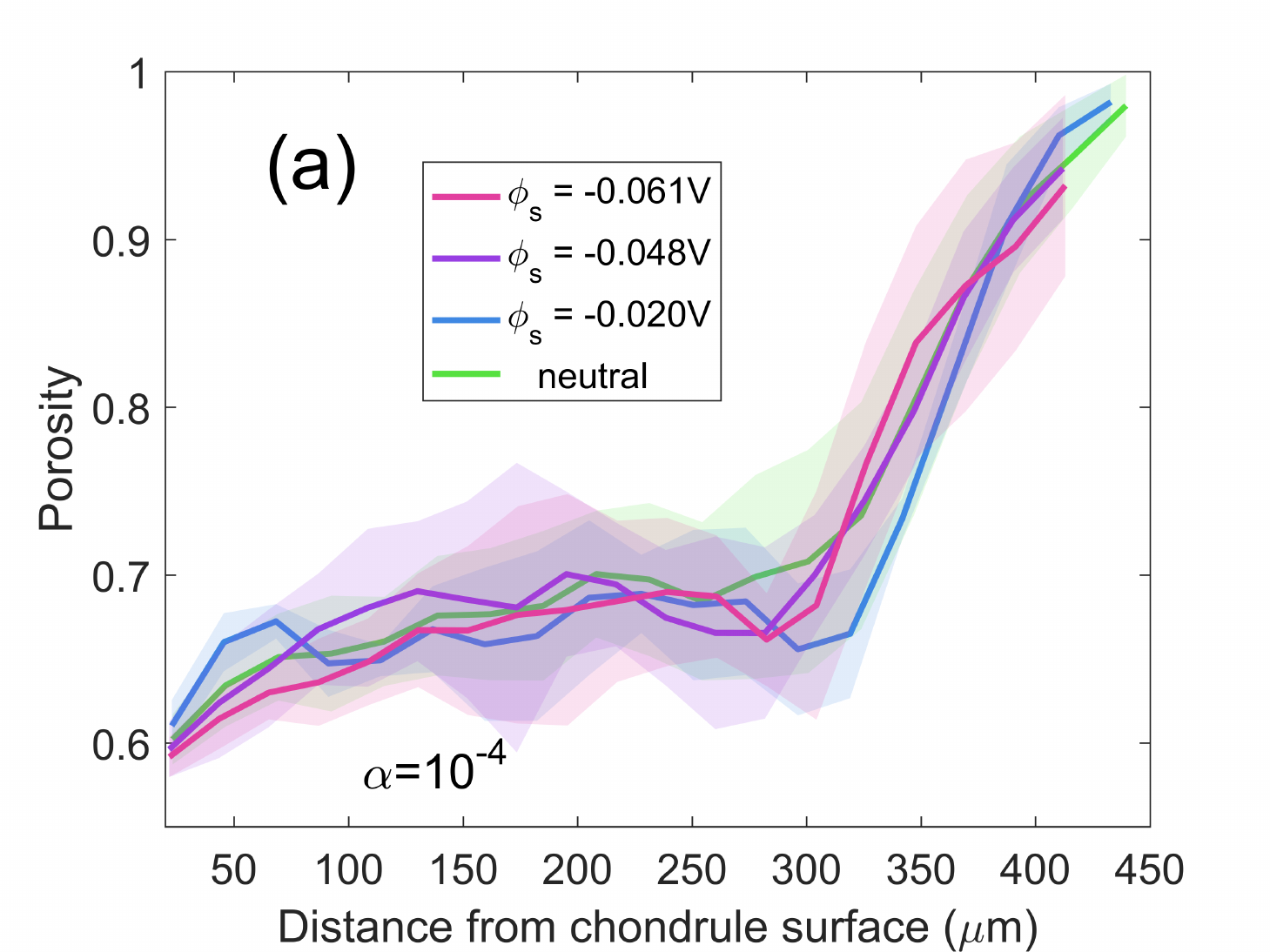}\includegraphics[width=6cm]{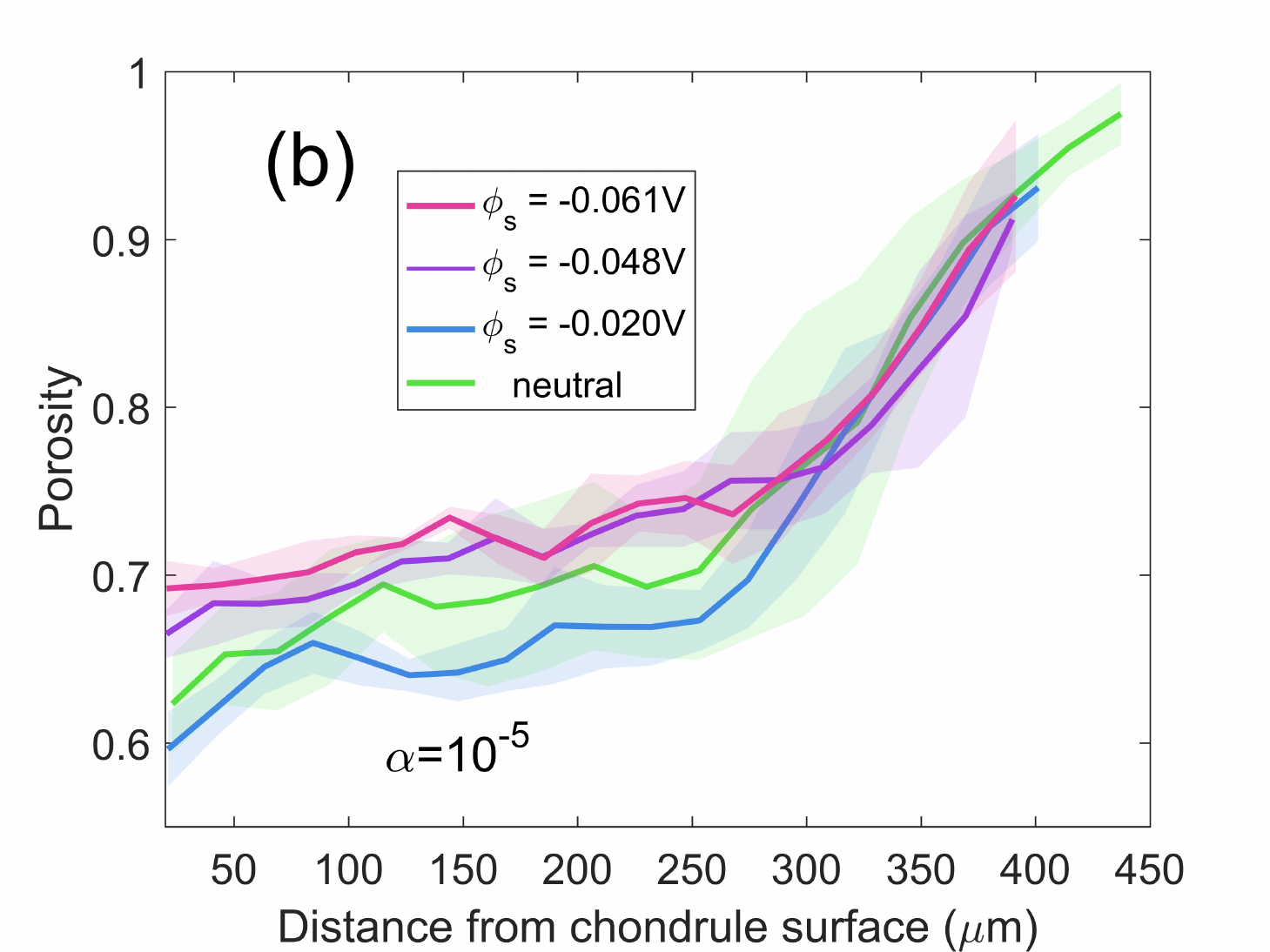}\includegraphics[width=6cm]{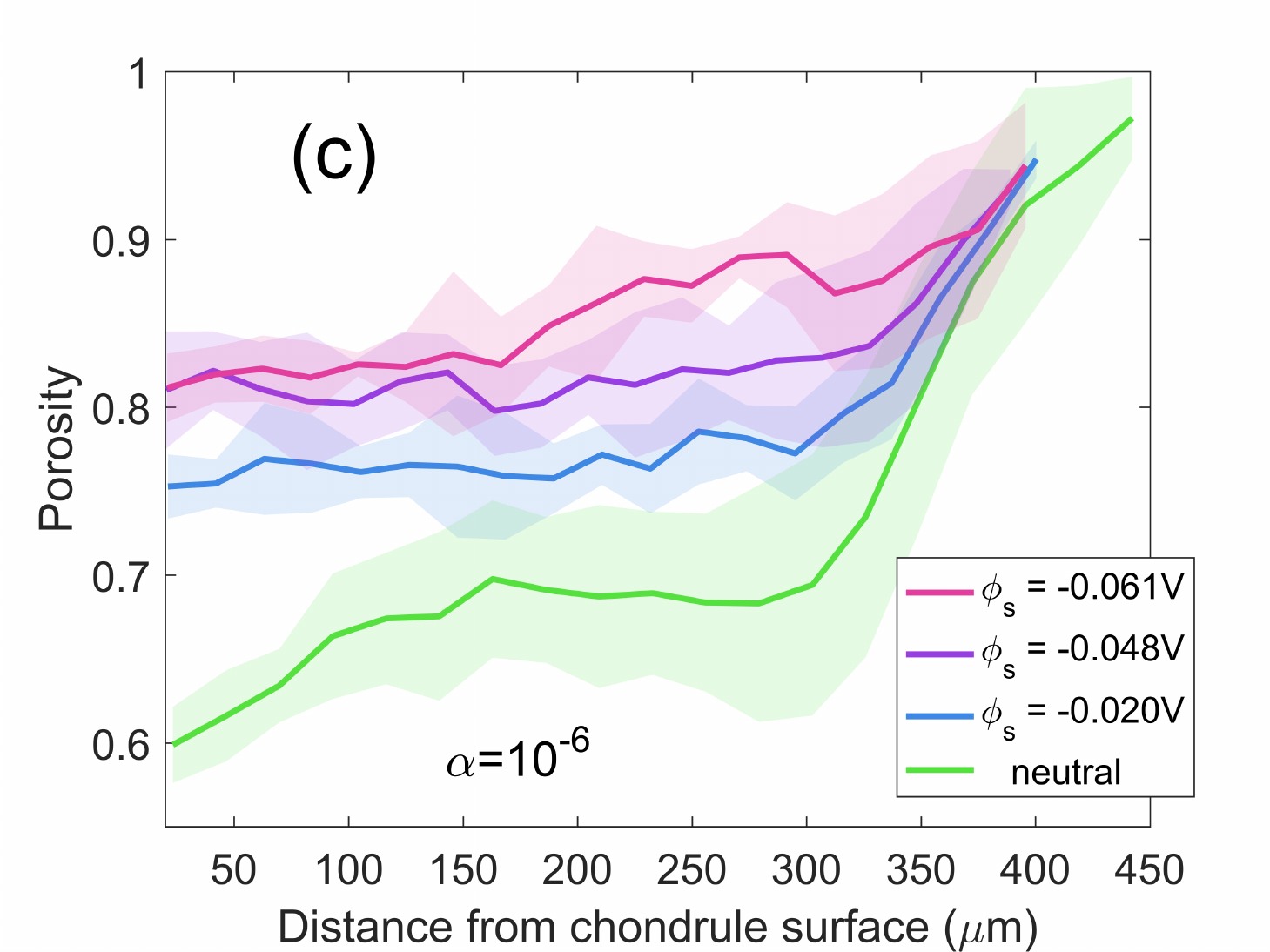}   
\caption{Radial profiles of dust rim porosity in each horizontal layer on a chondrule with a 700-$\mu$m radius, formed in neutral and charged environments, for equal rim thickness, defined by the distance from the chondrule surface encompassing 95\% of the total rim mass. The turbulence strength is a) $\alpha=10^{-4}$ , b) $\alpha=10^{-5}$, and c) $\alpha=10^{-6}$. Lines are averages for five chondrule rims, with the shaded area indicating the standard deviation.}
\label{f4}
\end{figure*}

In Fig. \ref{f4}, the change in rim porosity as a function of distance from the chondrule surface, for different charging conditions and turbulence strengths, is illustrated for a chondrule with radius $a=700\ \mu m$. Overall, the porosity increases from the base of the rim to the top, due to the grain-size coarsening toward outer portions of FGRs, consistent with observations. This is apparently caused by the fact that small grains pass through voids and fill in the lower rim layers (Metzler et al. 1992, Zega \& Buseck, 2003; see Paper 1). In relatively strong turbulence ($\alpha=10^{-4}$; Fig. \ref{f4}a), the charged and neutral dust rims have similar radial profiles for porosity. As turbulence weakens ($\alpha=10^{-5}$, $10^{-6}$; Fig. \ref{f4}b, c), the relative difference in the porosity between bottom and top layers decreases with increasing dust surface potential, because it is more difficult for large dust particles (which are able to overcome the Coulomb repulsion barrier) to pass through voids.

\begin{figure*}[!htb]
\includegraphics[width=6cm]{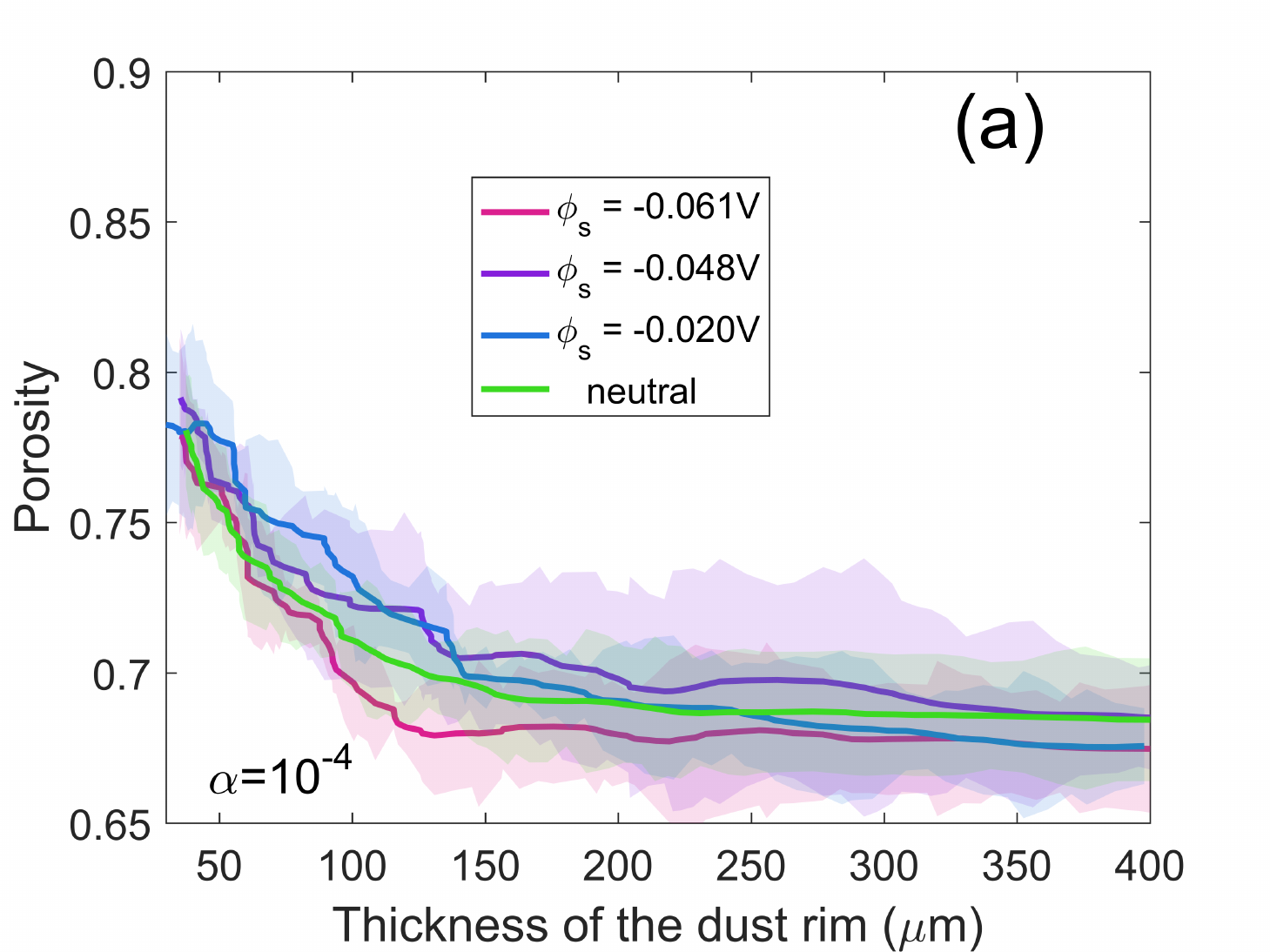}\includegraphics[width=6cm]{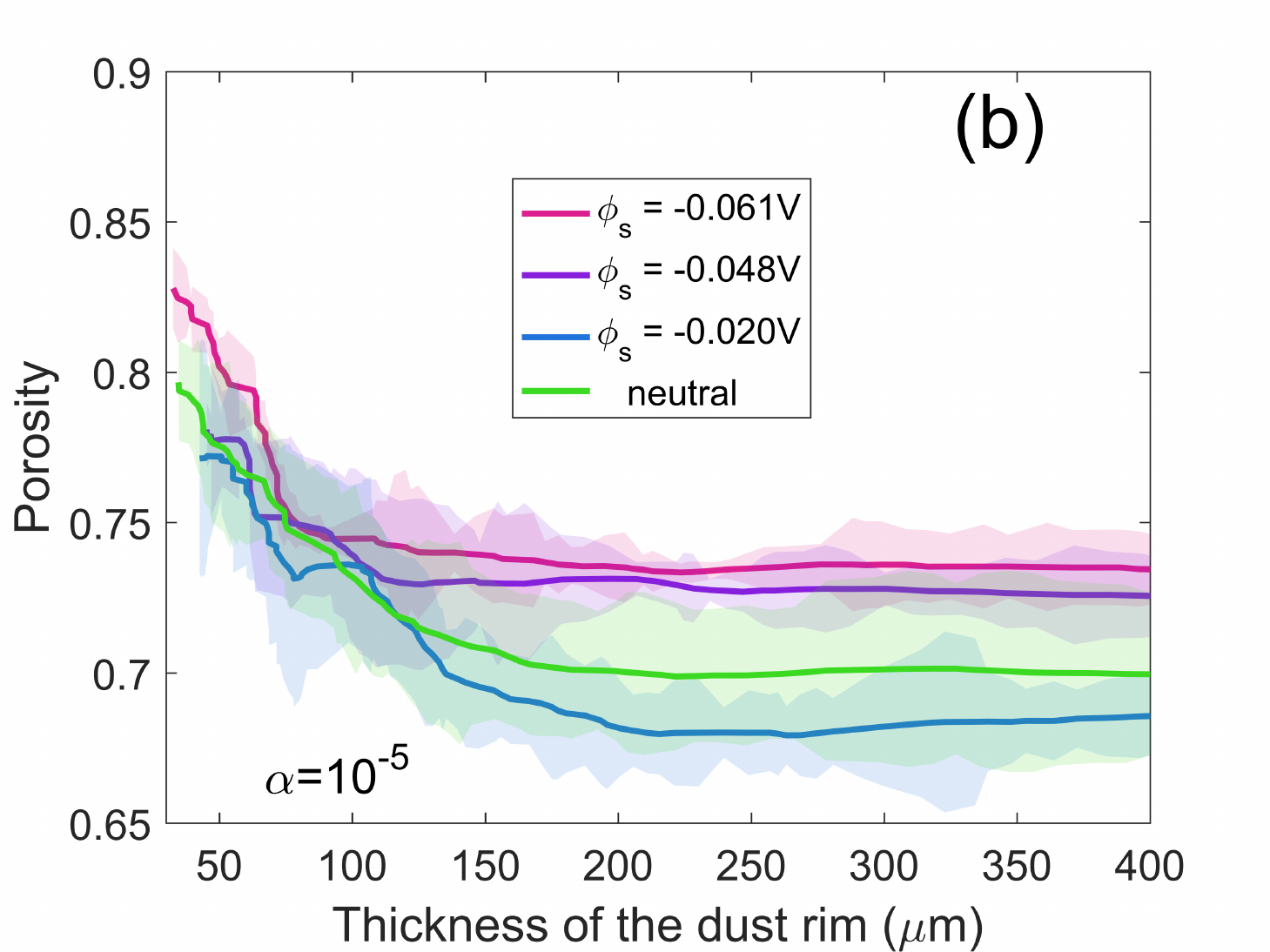}\includegraphics[width=6cm]{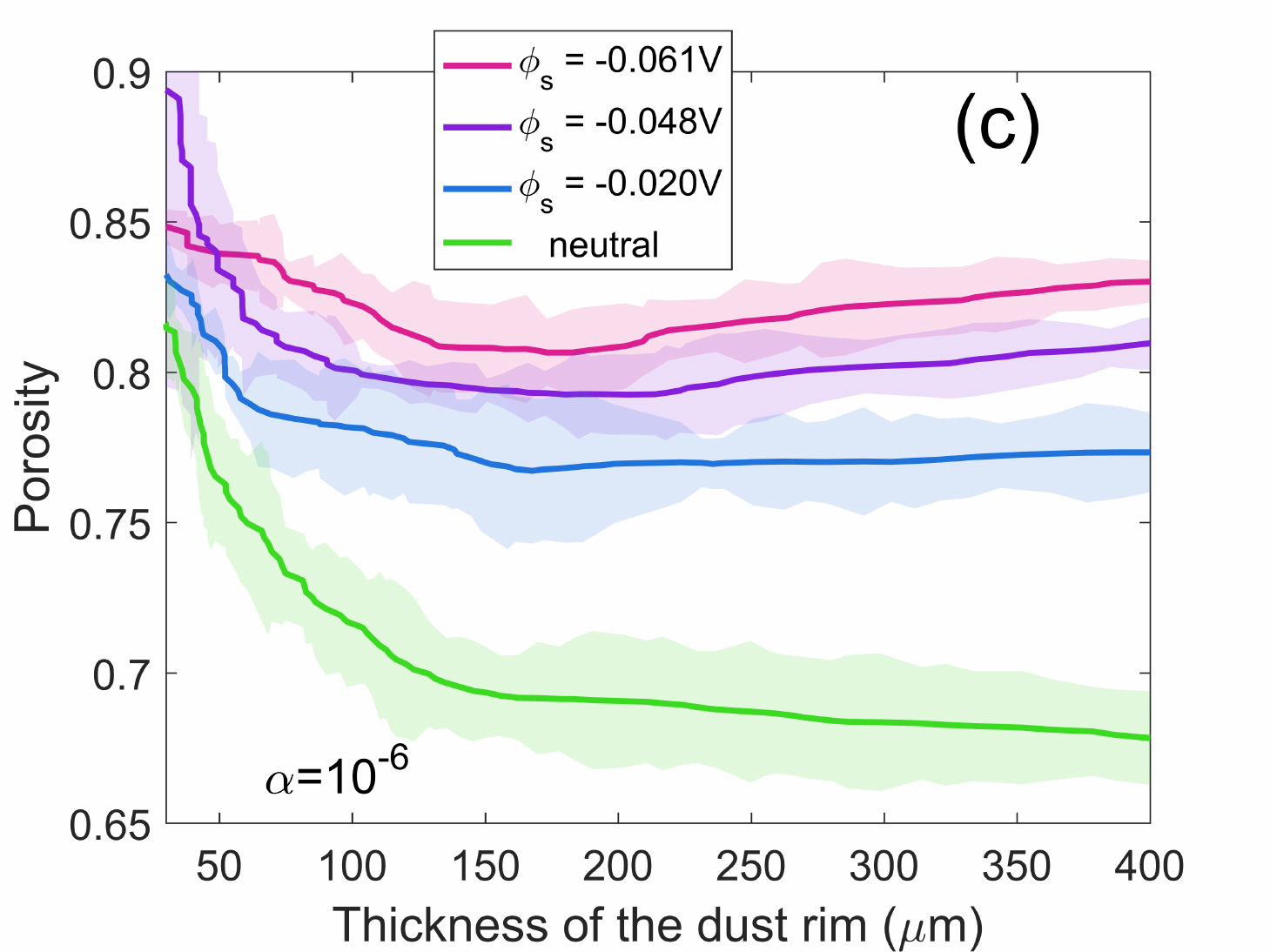}  
\caption{Change in average rim porosity as the rim is accumulated in different plasma conditions. Turbulence strength is a) $\alpha=10^{-4}$ , b) $\alpha=10^{-5}$, and c) $\alpha=10^{-6}$. Chondrule radius is 700 $\mu$m. Lines are averages for five chondrule rims, with the shaded area showing the standard deviation.}
\label{f2}
\end{figure*}

\begin{figure*}[!htb]
\includegraphics[width=9cm]{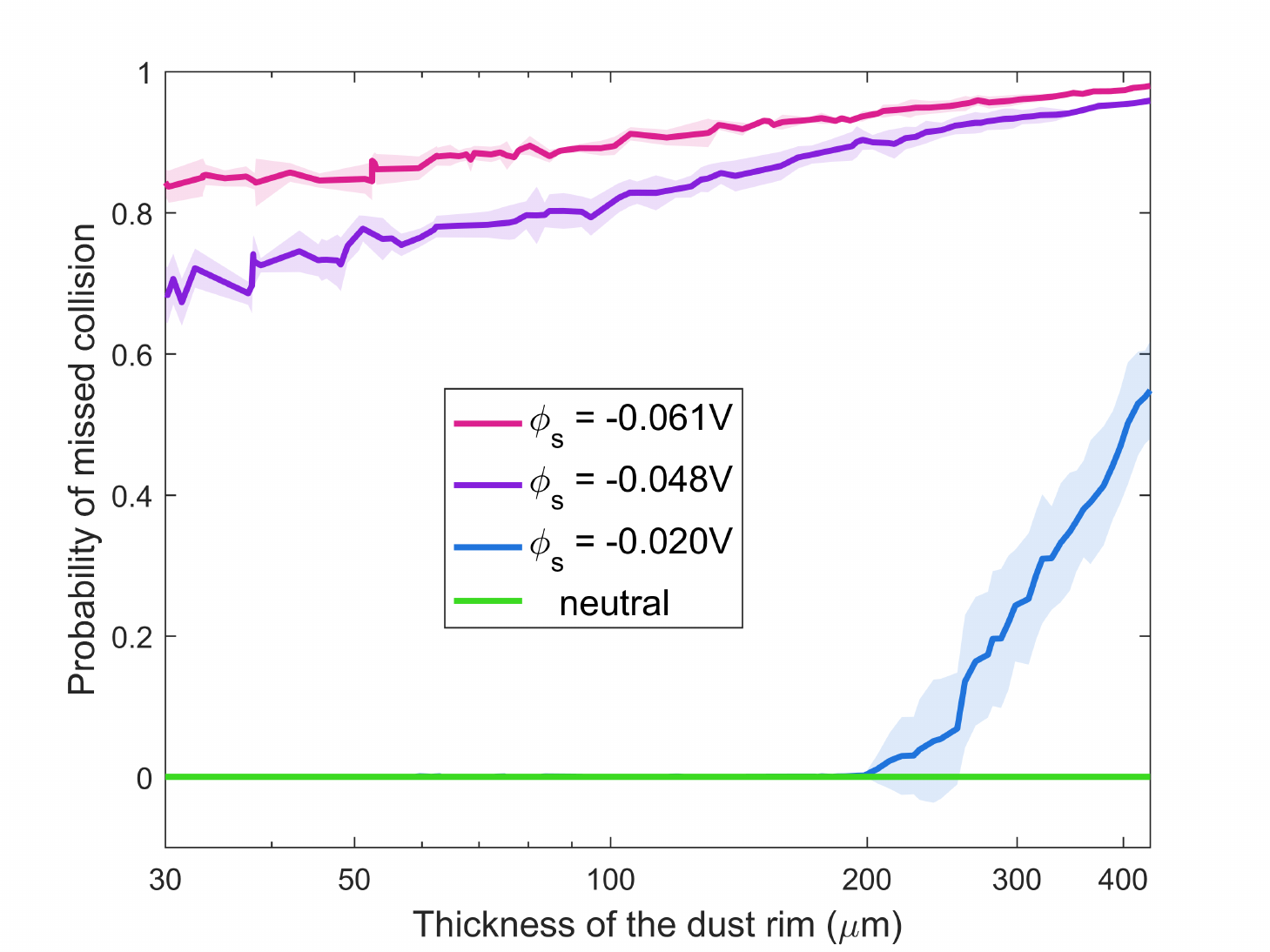} 
\caption{Probability of missed collision as the rim is accumulated in different plasma conditions, with turbulence strength $\alpha=10^{-5}$. Chondrule radius is 500 $\mu$m. Lines are averages for five chondrule rims, with the shaded area showing the standard deviation.}
\label{f3}
\end{figure*}

The evolution of the overall porosity of dust rims as the rims grow in thickness (defined as the average porosity of inner rim portion encompassing 95\% of the total rim mass), is shown in Fig. \ref{f2}. In strong turbulence with $\alpha \geqslant  10^{-4}$, the charged rims and neutral rims have similar porosity, which decreases as the rim is accumulated because the small colliding dust particles constantly fill in the gaps of the existing rim (an example of $\alpha = 10^{-4}$ is shown in Figure \ref{f2}a). In weak turbulence ($\alpha=10^{-6}$; Figure \ref{f2}c), higher charge increases the porosity of dust rims, and the porosity increases further as the rim grows. This is due to the increased likelihood of missed collisions with small grains caused by electrostatic repulsion, as shown in Figure \ref{f3}. The threshold size of dust particles which are able to overcome the Coulomb repulsion barrier increases as chondrules grow larger, and the lack of small monomers filling in the pore spaces results in higher porosity. The turbulence level $\alpha=10^{-5}$ is a turning point where the highly charged rims ($\phi _{s}$ = -0.061 V, -0.048 V) are more porous than the neutral rims and the weakly charged rims ($\phi _{s}$ = -0.020 V) are more compact than the neutral rims (Figure \ref{f2}b).


\begin{figure*}[!htb]
\includegraphics[width=9cm]{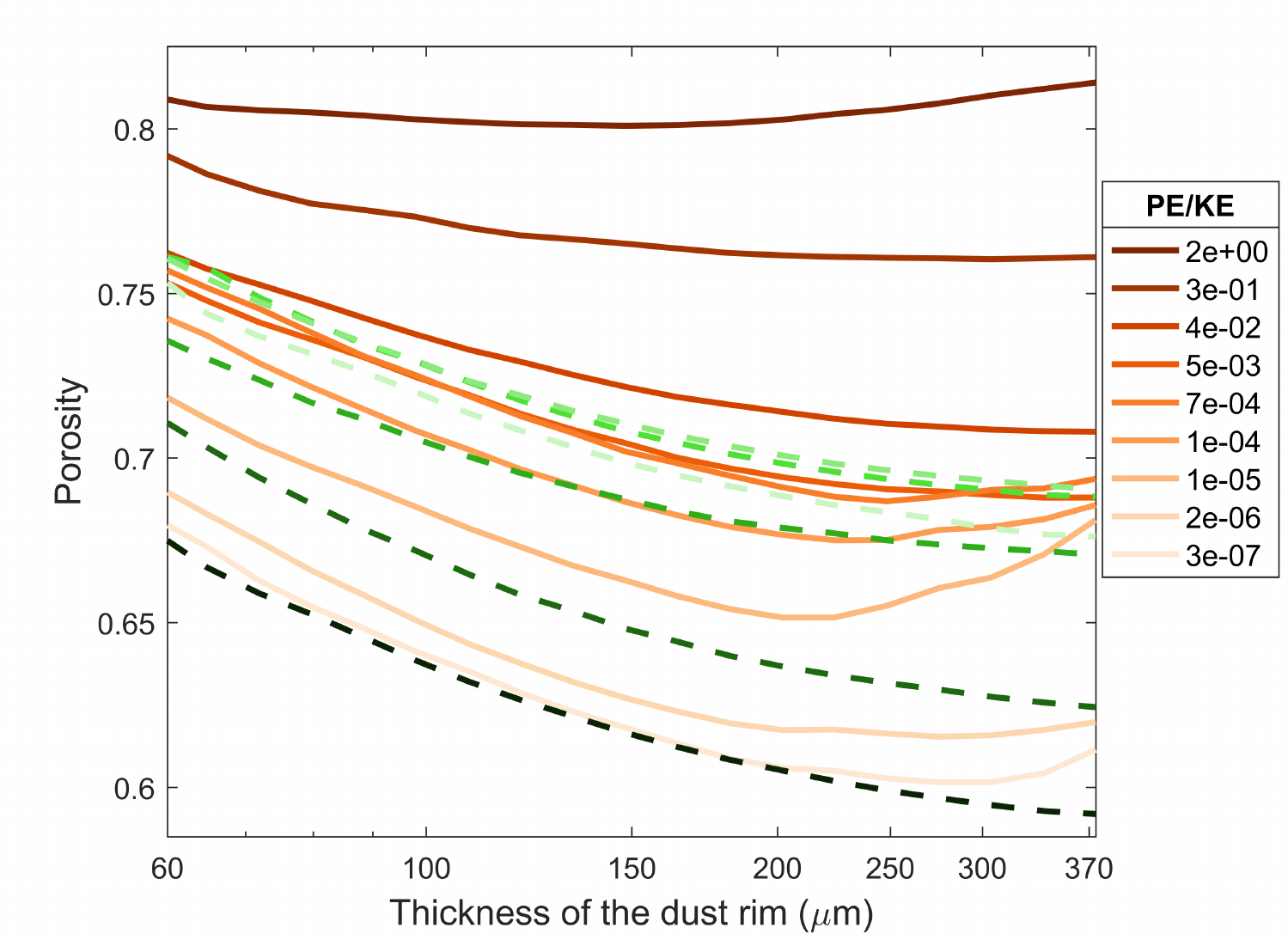} 
\caption{Change in average rim porosity as the rim is accumulated for different average values of PE/KE. Dust rims on chondrules of different sizes (a = 500-1000 $\mu$m, in 100-$\mu$m increments),  formed in different plasma conditions ($\phi _{s}$ = -0.061 V, -0.048 V, -0.020 V), with different turbulent strengths ($\alpha=10^{-1}$, $10^{-2}$, $10^{-3}$, $10^{-4}$, $10^{-5}$, $10^{-6}$) are binned into nine groups based on the average PE/KE. Shown is the average porosity in each group.  For comparison, the average porosity of neutral rims for different turbulence levels are shown in green ($\alpha=10^{-1}-10^{-6}$ in order of decreasing color shades).}
\label{f5}
\end{figure*}

In Fig. \ref{f5}, the evolution of rim porosity as it is accumulated is summarized for all combinations of turbulence strengths, dust surface potentials and chondrule sizes, characterized by the average PE/KE for the chondrule-dust interactions for a given set of parameters. The average porosity of dust rims decreases as they grow in thickness over time, except for values of PE/KE $\gtrsim$ 2, which result in a slight increase of porosity with time. Rims with lower PE/KE have more compact structure and the porosity decreases more rapidly than those with higher PE/KE. [Note that PE/KE is the average calculated for all dust particle sizes, so even if PE/KE $> 1$ for the dust population, the largest dust particles in the simulation may have enough energy to overcome the Coulomb repulsion barrier; PE/KE is mainly determined by the size of the chondrule core for given environmental conditions, and only changes a little as the rim is accumulated].   

\begin{figure*}[!htb]
\includegraphics[width=9cm]{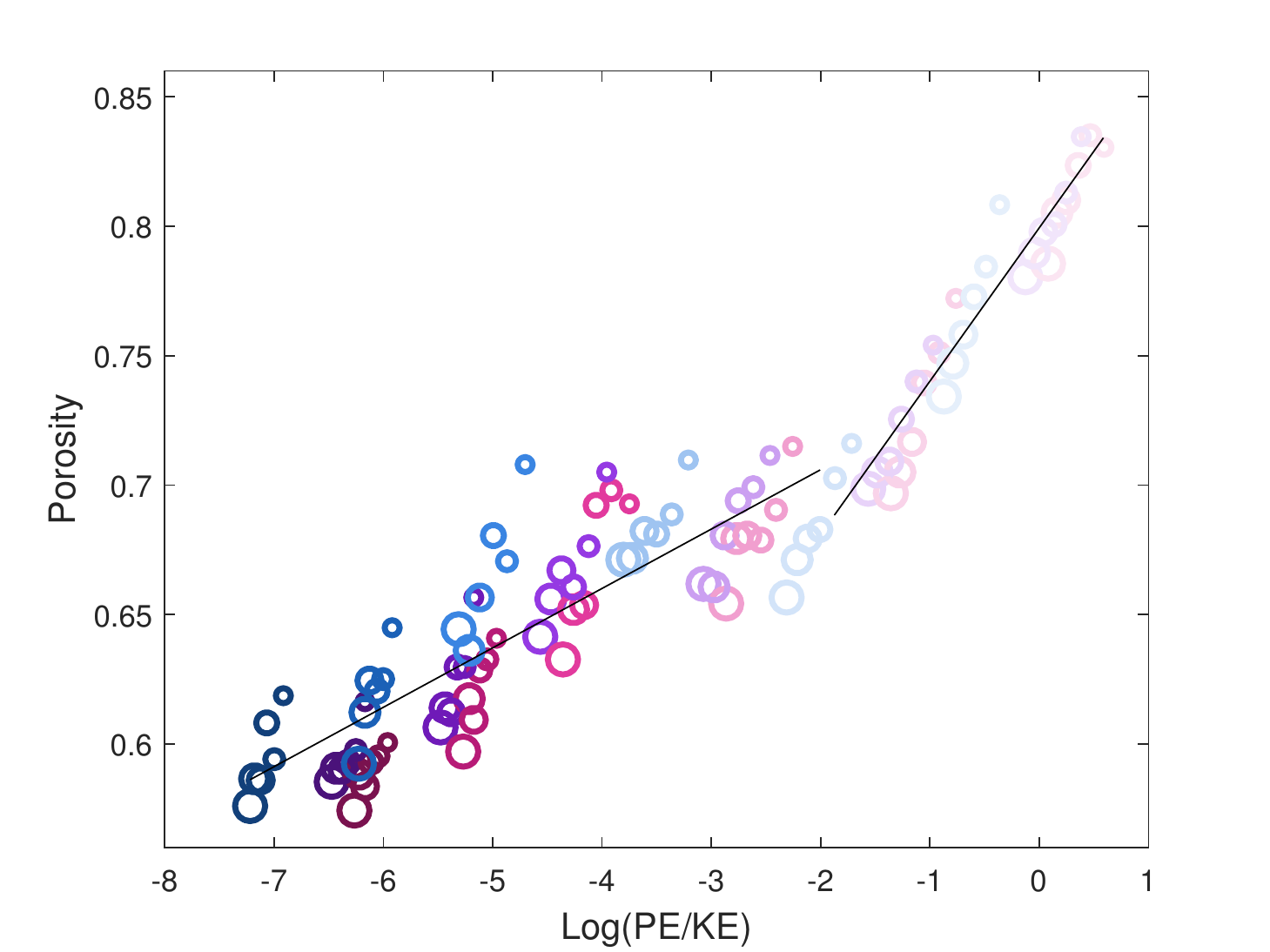} 
\caption{Comparisons of porosity of dust rims with a thickness of 300 $\mu$m as a function of the ratio of grains’ electrostatic potential energy to the kinetic energy. Dust surface potential is indicated by color (pink: $\phi _{s}$ = -0.061V; purple: $\phi _{s}$ = -0.048V; blue: $\phi _{s}$ = -0.020V). Turbulence level is denoted by shade ($\alpha=10^{-1}$ to $\alpha=10^{-6}$ in order of decreasing color shades). The size of the chondrule core is represented by symbol size (r = 500-1000 $\mu$m, in 100 $\mu$m increments). The black lines are the linear, least-square polynomial fit to the data points.}
\label{f6}
\end{figure*}

Fig. \ref{f6} compares the average porosity of the entire rim when the rims have reached a thickness of 300 $\mu$m for chondrules in environment with different PE/KE. The porosity $\psi$ is almost linearly proportional to the logarithm of PE/KE. Two fit lines are shown: $\psi  \approx 0.023 \log(PE/KE)+0.75$ for $log(PE/KE) \lesssim -2$, and $\psi  \approx 0.060 \log(PE/KE)+0.80$ for $log(PE/KE) \gtrsim -2$. The greater slope of the fit line for large PE/KE indicates a greater dependence of the porosity on the charge in weak turbulence. Given the same environment, larger chondrules tend to accumulate less porous rims than small chondrules, due to the greater relative velocity with respect to dust particles, which increases the restructuring and reduces the repulsion of small particles.

\subsection{Time to accumulate rims}

\begin{figure*}
\includegraphics[width=9cm]{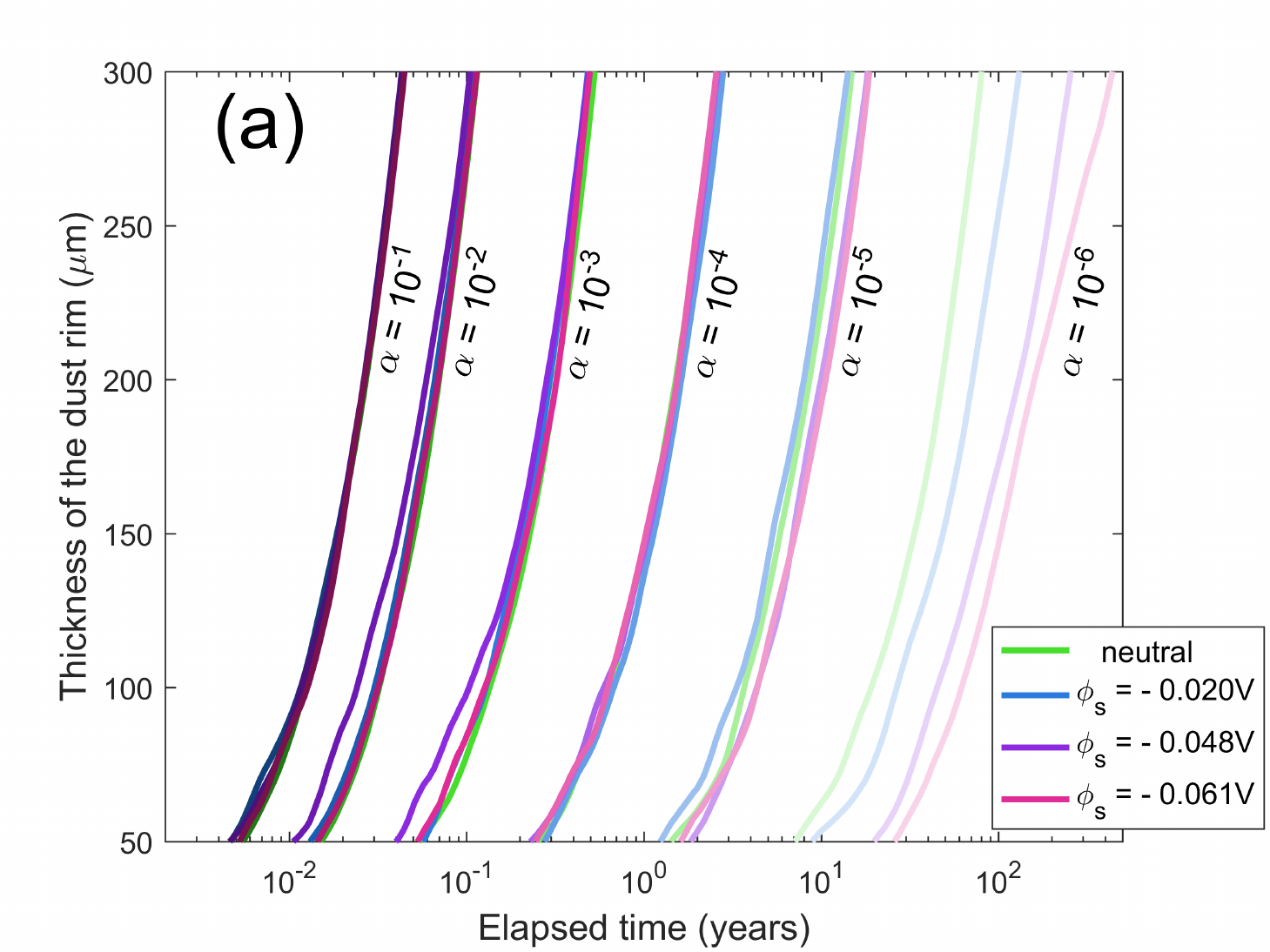}\includegraphics[width=9cm]{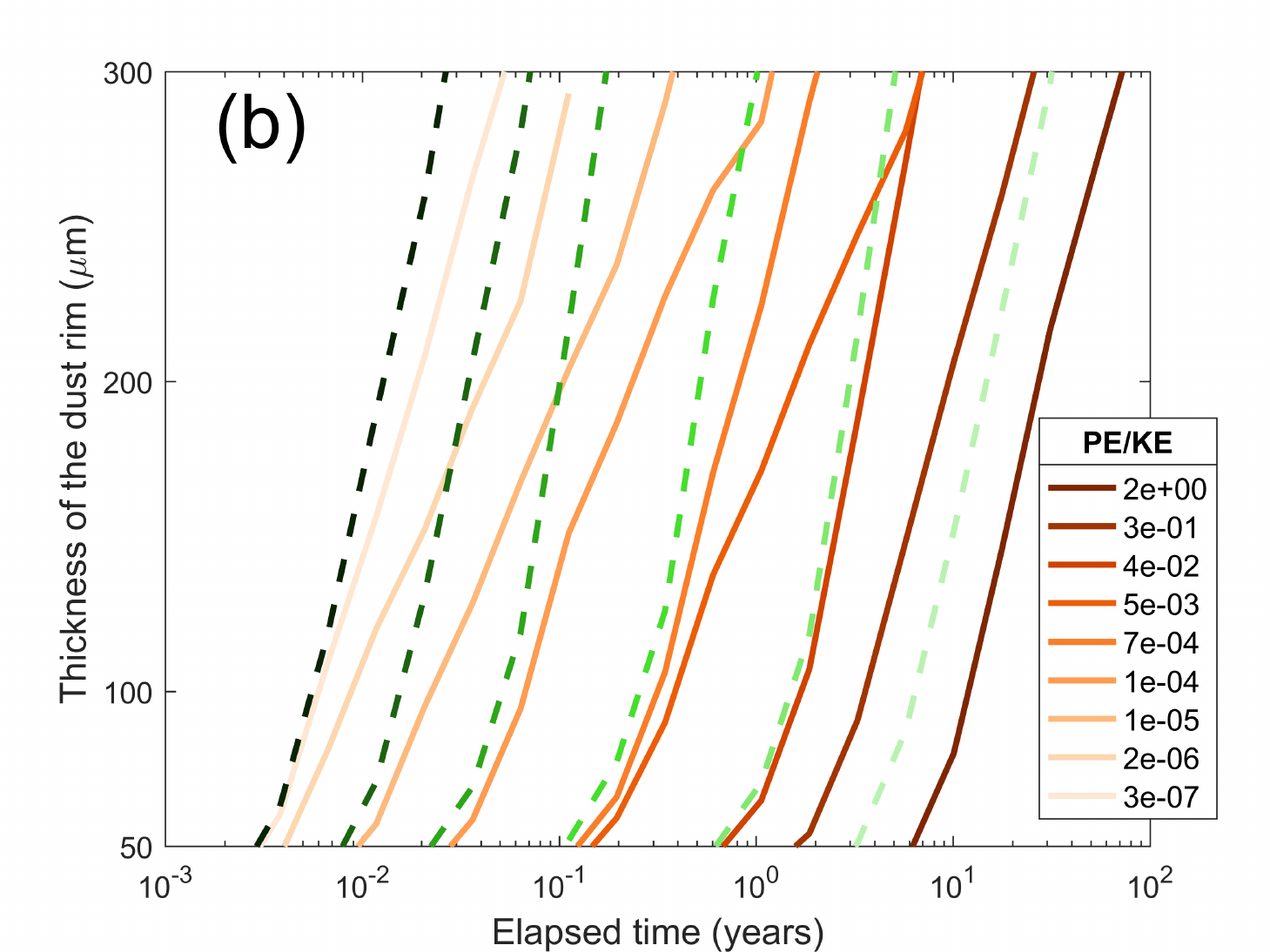}
\caption{a) Thickness of the dust rim on a 500-$\mu$m-radius chondrule, as a function of elapsed time, in charged (pink: $\phi _{s}$ = -0.061V; purple: $\phi _{s}$ = -0.048V; blue: $\phi _{s}$ = -0.020V) and neutral (green lines) environments, with different turbulence strengths ($\alpha=10^{-1}$ to $\alpha=10^{-6}$ in order of decreasing color shades). b) Thickness of the dust rim as a function of elapsed time for different average values of PE/KE. Dust rims on chondrules are binned into 9 groups based on the average PE/KE. The green dashed lines indicate the growth of neutral dust rims in environments with different turbulence levels ($\alpha=10^{-1} - 10^{-6}$ in order of decreasing color shades).}
\label{f10}
\end{figure*}

As shown in Paper I, in a neutral environment, chondrules in strong turbulence accrete dust rims faster than those of the same size in weak turbulence, and large chondrules accumulate dust rims faster than small chondrules in the same environment. Here we examine the effect of charge on the time it takes a chondrule to collect a dust rim. Although the growth rates of charged and neutral rims are similar in strong turbulence, they can differ markedly in weak turbulence, as illustrated in Fig. \ref{f10}a for a 500-$\mu$m chondrule. Differences are seen among the charged and neutral rims for turbulence levels $\alpha \leqslant 10^{-5}$, due to a large fraction of small dust particles being repelled. The critical value of turbulence at which the growth rate of charged rims starts lagging behind neutral rims differs for different chondrule sizes: it shifts towards weaker turbulence with increasing chondrule size due to the greater kinetic energy of dust particles. More complete data for all chondrule sizes, turbulence levels and charge levels are presented in Fig. \ref{f10}b, where the growth rate is shown as a function of PE/KE. Overall, the higher the charge and the weaker the turbulence, i.e., greater PE/KE, the more the growth rates of charged rims lag behind those of neutral rims.

\begin{figure*}[!htb]
\includegraphics[width=9cm]{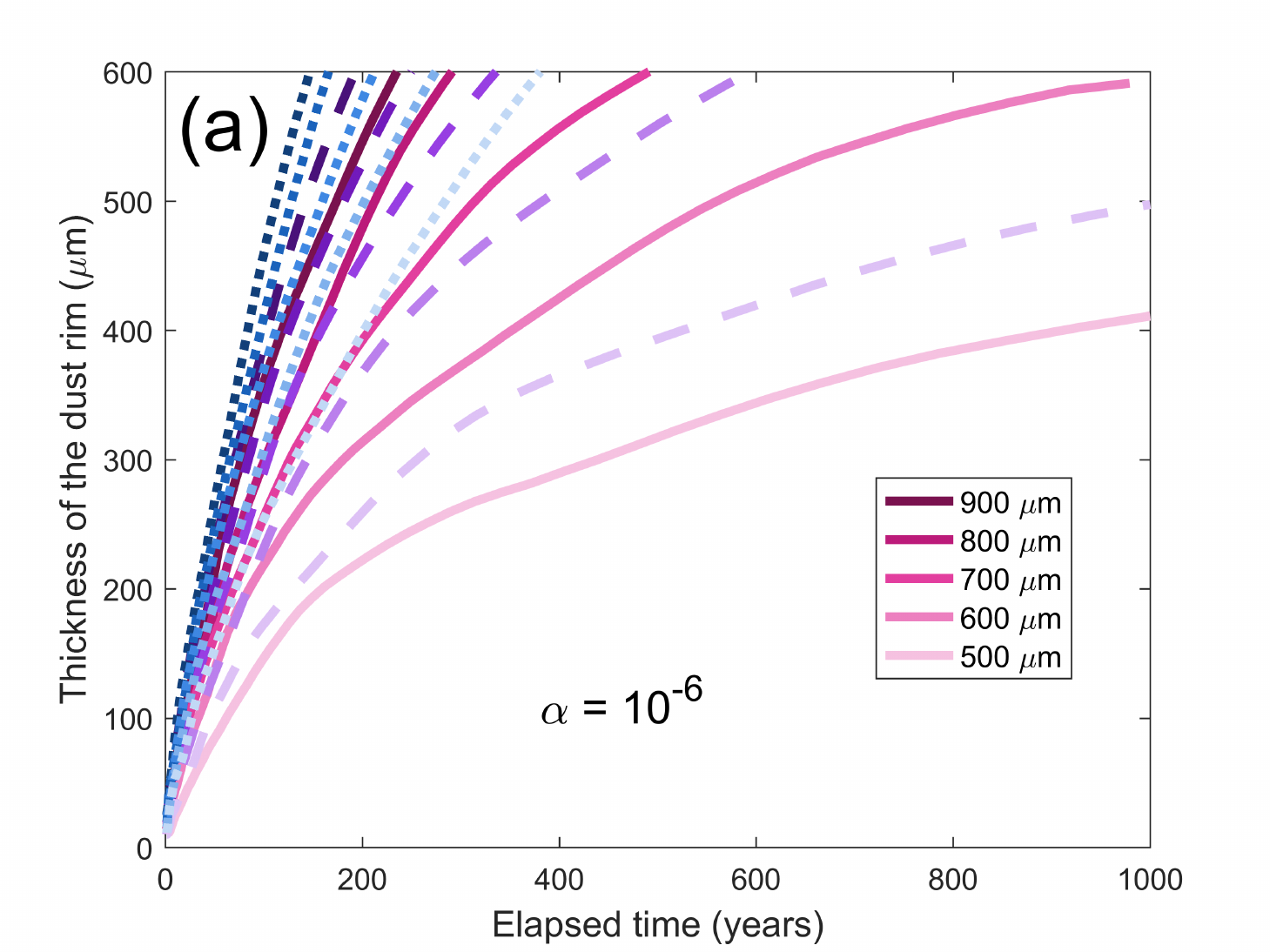}\includegraphics[width=9cm]{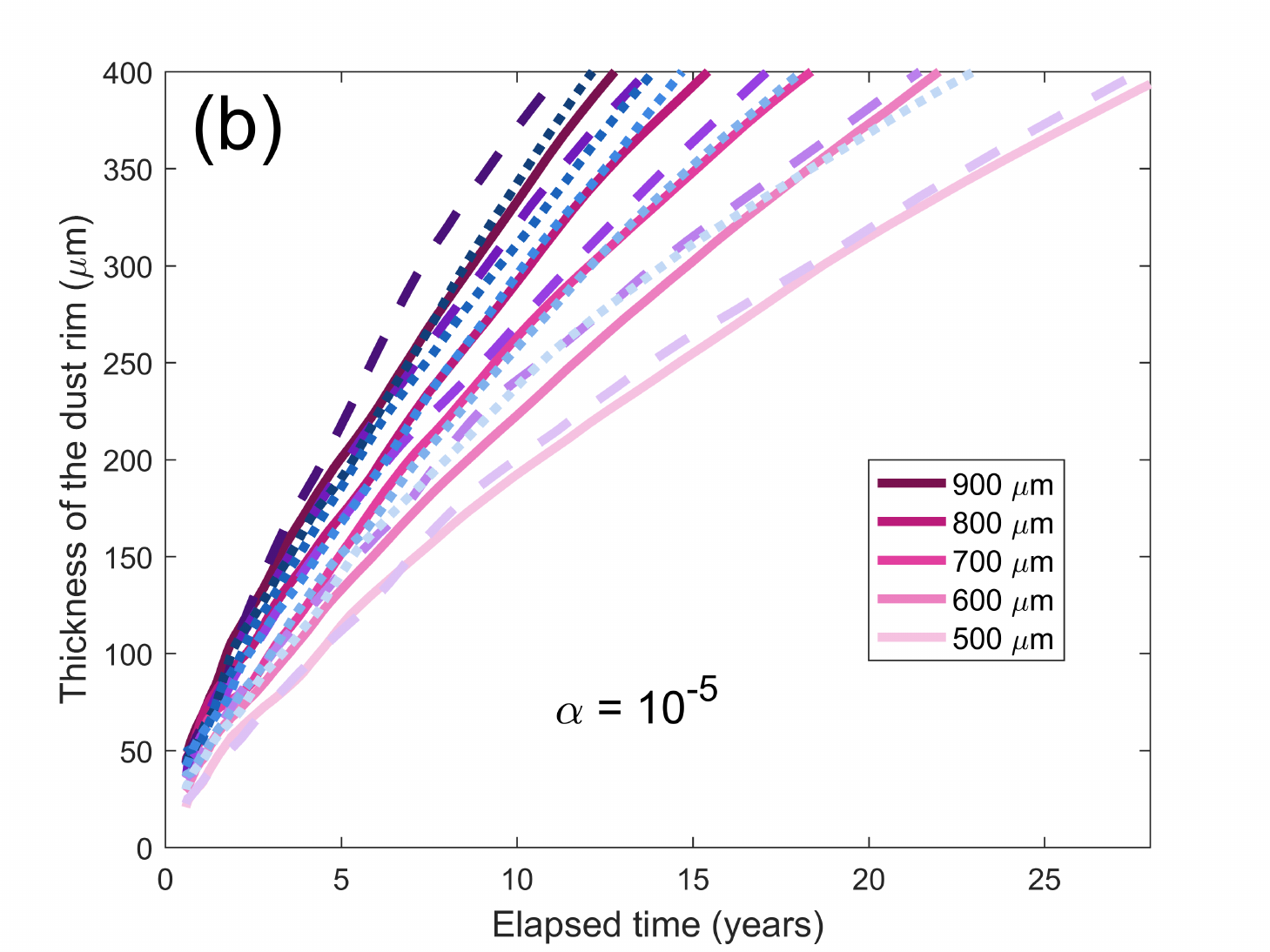} 
\caption{Thickness of the dust rim on chondrules with different radii as a function of elapsed time, for different plasma conditions (pink solid lines: $\phi _{s}$ = -0.061V; purple dashed lines: $\phi _{s}$ = -0.048V; blue dotted lines: $\phi _{s}$ = -0.020V). Turbulence strength is a) $\alpha=10^{-6}$ and b) $\alpha=10^{-5}$}.
\label{f12}
\end{figure*}

Figure \ref{f12} compares the growth rates of chondrules of different sizes in more detail. In the weakest turbulence ($\alpha=10^{-6}$), it is shown that the rim ceases accretion before it grows to a thickness of 400 $\mu$m for the smallest chondrule (500 $\mu$m) with highest surface potential ($\phi _{s}$ = -0.061V). Therefore, in low turbulence, the presence of charge not only slows the growth rate, but can cause the rim growth to stop. The maximum rim thickness depends not only on the turbulence and charge level, but also on the chondrule size. Although for the range of the conditions and elapsed times examined here, rim growth is halted only for the smallest chondrule, larger chondrules are also expected to eventually stop accreting dust rims after reaching some maximum thickness depending on the surface charge and turbulence level.

\section{Discussion and conclusions}



We have compared the growth of chondrule rim in neutral and weakly ionized gas in PPDs, where the collisions leading to rim growth are driven by turbulence.  The turbulence strength, the amount of charge, and the chondrule size all come together to determine the ratio of a dust particle's potential energy at the point of impact to its kinetic energy far from the chondrule. Thus differences in the porosity, thickness, and the distribution of dust sizes within the rim, as well the time needed to accumulate a rim of a given thickness, are shown to depend on the ratio of PE/KE.  In general, greater charge, weaker turbulence, and smaller chondrule size results in larger average monomer size, increased porosity, smaller rim thickness and greater time to accumulate the rim.


The radial profiles of dust rim porosity show that the outer regions have a higher porosity than the inner regions in all cases, and the values vary in different environments. In general, dust rims formed in stronger turbulence are more compact than those formed in weak turbulence due to more severe restructuring (see Paper I) and more small dust grains filling the pores. The impact of the charge on the rim porosity varies in different turbulence regimes. In a strongly turbulent environment ($\alpha \geqslant  10^{-4}$), the charge can either increase or decrease the rim porosity due to the two factors: first, the reduced relative velocity caused by the electrostatic force reduces restructuring which increases the porosity; second, the electrostatic force can alter the particle trajectories as they pass through the gaps in the rim, and the particles tend to avoid the extremities of the rim to minimize the potential energy of the configuration, which causes a more compact arrangement of monomers and decreases the porosity [note that this effect is less important when particles move fast]. The impact of the charge is negligible for $\alpha \geqslant  10^{-2}$, and FGR porosity has an approximately constant value of 50 -- 65\% in the inner regions of the rims formed in both charged and neutral environments. For medium turbulence levels ($\alpha=10^{-3}$, $10^{-4}$), charged rims are overall slightly more compact than neutral rims, but they both fluctuate in the range of 60 -- 70\%. Modelling and laboratory estimates (Ormel et al. 2008; Dominik et al. 1997; Blum 2004) show that the rolling motion within the dust layer or restructuring caused by greater collisional energy can increase the filling factor ($\phi_{\sigma }=1-\psi $) to a value of $\sim$ 0.33, which is close to our results. In the case of weak turbulence ($\alpha \lesssim 10^{-5}$), the presence of charge on the dust results in the very smallest dust particles being repelled from the chondrule surface. This lack of small dust grains leads to an increase in the porosity of the dust rims, as the small dust grains tend to fill in the pore spaces. Greater values of PE/KE result in more small grains being repelled and therefore higher rim porosity, and the difference between rims with different PE/KE increases as the rim grows (Fig. \ref{f5}). The inner regions of charged rims have a porosity of 60 -- 75\% for $\alpha=10^{-5}$ and 70 -- 92\% for $\alpha=10^{-6}$, while the porosity of neutral dust rims ranges from 63\% to 72\% for these turbulence strengths. 

Note that these are the initial porosities that dust rims acquired in the PPD, which can be greatly reduced during following compaction process caused by low-intensity shocks that individual rimmed chondrules or aggregates of chondrules experience before incorporated into parent bodies (strong shocks can melt chondrules; Desch et al. 2012; Thompson 1985) or by energetic collisions between agglomerations of rimmed chondrules (Bland 2011). Bland et al. (2011) researched the relationship between fabric intensity and net compression by examining the degree of alignment of olivine grain in FGRs, which indicates the amount of deformation, and reconstructed an initial rim porosity of 70--80\%. Our results have a broader range of porosity due to the variety of conditions considered. In addition to the porosity, the repulsion of small dust grains also affects the monomer size distribution within rims, and the average monomer size is positively related to PE/KE for $\alpha \gtrsim 10^{-2}$ (Fig. \ref{f7}).

\begin{figure*}[!htb]
\includegraphics[width=9cm]{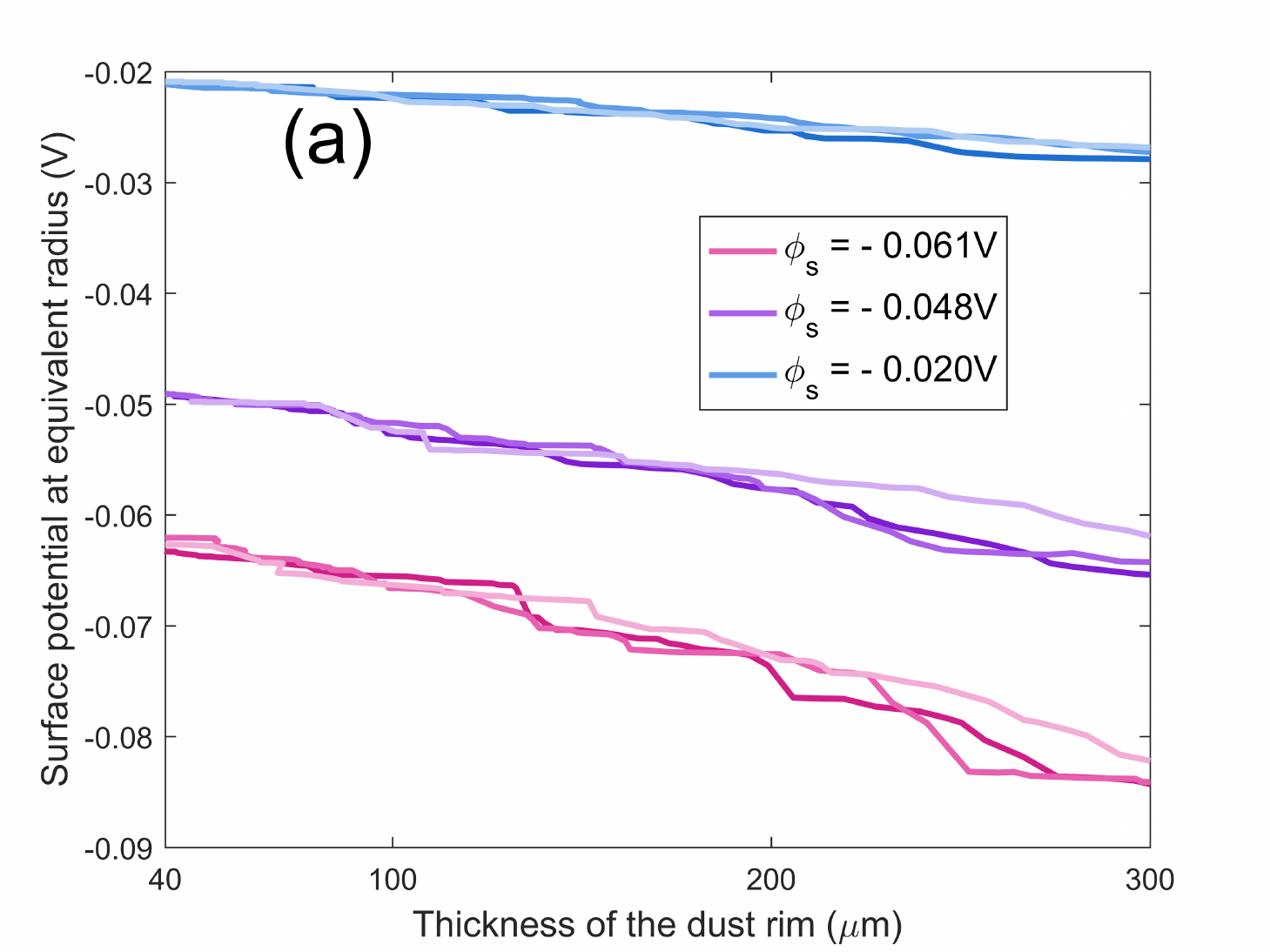}\includegraphics[width=9cm]{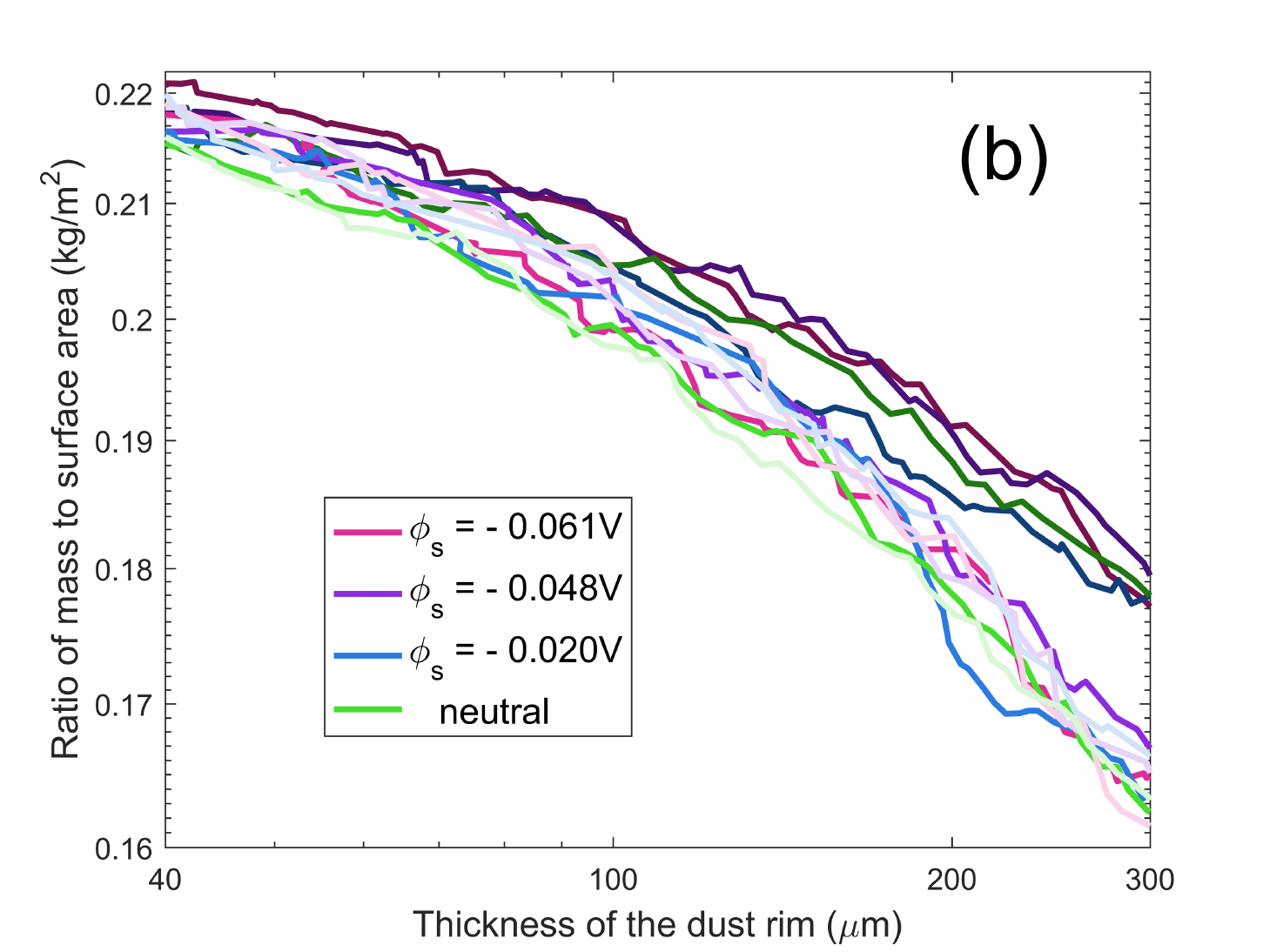} 
\caption{a) Charge and b) mass to surface ratio of rimmed chondrule with a radius of 700 $\mu$m as the rim is accumulated, for different dust surface potentials (pink: $\phi _{s}$ = -0.061V; purple: $\phi _{s}$ = -0.048V; blue: $\phi _{s}$ = -0.020V), and different turbulence levels ($\alpha=10^{-1}$, $10^{-3}$ and $10^{-5}$ in order of decreasing color shades. }
\label{f16}
\end{figure*}

FGRs formed in environments with strong turbulence grow more rapidly than those in weak turbulence (Ormel et al. 2008; Paper I). In strong turbulence, the charge affects the formation time of dust rims by changing the rim porosity and equivalent radius of the dust rim, which results in a different surface-to-mass ratio and thus relative velocity with respect to dust particles (see Eq. \ref{eq:vturb}). In low turbulence, the charge impacts the rim porosity and formation time mainly through the repulsion of small dust particles. As chondrules grow in size and rims become more porous, the increased surface area results in a higher surface potential, as shown in Fig. \ref{f16}a, which poses a greater electrostatic barrier for dust particles. Meanwhile, the ratio of mass to surface area of a rimmed chondrule decreases as the thickness of the porous rim relative to the compact chondrule core increases (Fig. \ref{f16}b), which decreases the relative velocity between chondrule and dust particles. Both factors cause more small dust particles to be repelled. Therefore, the rim becomes more porous and the growth rate slows down over time. Although larger chondrules have a higher surface potential than small chondrules, they also have a greater mass-to-surface area ratio increasing the relative velocity, which enables the dust particles to overcome the increased electrostatic barrier. Therefore, large chondrules grow faster, and form thicker and more compact rims than small chondrules. In very weak turbulence, the presence of charge not only slows the growth rate, but can cause the rim growth to stop. The maximum thickness depends on PE/KE: the lower the turbulence, the higher the charge, and the smaller the chondrule size, the thinner the dust rim that can be formed.

The cessation of rim growth caused by the charge is similar to the charge barrier for aggregate growth identified by Okuzumi (2009). However, there are several mechanisms that could prevent the freezing of rim growth. First, chondrules may resume accreting dust after traveling to environments with less charge or stronger turbulence (through radial infall, etc.). A sudden change in porosity or monomer size distribution within the rims may imply they were formed in multiple locations. Second, vertical mixing of dust particles can enhance the kinetic energy of dust particles: dust particles that couple strongly to the gas are lifted out of the region, and reenter it after they grow larger (Okuzumi 2011b). Third, positive charging of the dust grains caused by photoelectric emission due to stellar radiation (Akimkin 2015) can remove the electrostatic barrier. However, neither of the last two mechanisms is significant in dense regions of the disk (Ivlev 2016). Finally, compaction of dust rims caused by thermal alternation or nebular shock waves results in a higher mass-to-surface ratio which increases the relative velocity between chondrules and dust particles, and a lower charge-to-mass ratio which reduces the electrostatic barrier. Both factors promote the growth of dust rims.


Our main conclusions are: 

\begin{itemize}
 \item The overall porosity of neutral dust rims decreases as they grow in thickness, while that of charged rims can either decrease (low PE/KE) or increase (high PE/KE). In weakly turbulent regions, higher charge results in higher rim porosity, and this increases as the rim grows. 

 \item Deviation from the initial grain size distribution (with a greater proportion of large monomers incorporated into the rim) increases for greater charge, weaker turbulence, and smaller chondrule cores. In addition, the deviation from the initial monomer distribution increases as the rim increases in thickness (given more time to accumulate).  
 

 \item The higher the charge and the weaker the turbulence, the more the growth rates of charged rims lag behind those of neutral rims. In low turbulence ($\alpha < 10^{-4}$), the presence of charge not only slows the growth rate, but can cause the rim growth to stop.
\end{itemize}

\textbf{Acknowledgments:} Support from the National Science Foundation grant PHY-1707215 is gratefully acknowledged.

\appendix

\section*{Appendices}
\addcontentsline{toc}{section}{Appendices}
\renewcommand{\thesubsection}{\Alph{subsection}}




\def\bibindent{1em}

\end{document}